\documentclass[twocolumn,english,nofootinbib]{revtex4}
\usepackage[T1]{fontenc}
\usepackage[latin9]{inputenc}
\usepackage{color}
\usepackage{float}
\usepackage{amsmath}
\usepackage{graphicx}
\usepackage{amssymb}

\makeatletter

\providecommand{\tabularnewline}{\\}

\@ifundefined{textcolor}{}
{%
 \definecolor{BLACK}{gray}{0}
 \definecolor{WHITE}{gray}{1}
 \definecolor{RED}{rgb}{1,0,0}
 \definecolor{GREEN}{rgb}{0,1,0}
 \definecolor{BLUE}{rgb}{0,0,1}
 \definecolor{CYAN}{cmyk}{1,0,0,0}
 \definecolor{MAGENTA}{cmyk}{0,1,0,0}
 \definecolor{YELLOW}{cmyk}{0,0,1,0}
 }

\makeatother

\usepackage{babel}

\begin{document}

\title{Viscous damping of r-modes: Small amplitude instability}

\author{Mark G. Alford, Simin Mahmoodifar and Kai Schwenzer}

\address{Department of Physics, Washington University, St. Louis, Missouri,
63130, USA}
\begin{abstract}
We study the viscous damping of r-modes of compact stars and analyze
in detail the regions where small amplitude modes are unstable to
the emission of gravitational radiation. We present general expressions
for the viscous damping times for arbitrary forms of interacting dense
matter and derive general semi-analytic results for the boundary of
the instability region. These results show that many aspects, such
as the physically important minima of the instability boundary, are
surprisingly insensitive to detailed microscopic properties of the
considered form of matter. Our general expressions are applied to
the cases of hadronic stars, strange stars, and hybrid stars, and
we focus on equations of state that are compatible with the recent
measurement of a heavy compact star. We find that hybrid stars with
a sufficiently small core can {}``masquerade'' as neutron stars
and feature an instability region that is indistinguishable from that
of a neutron star, whereas neutron stars with a core density high
enough to allow direct Urca reactions feature a notch on the right
side of the instability region.
\end{abstract}
\maketitle

\section{Introduction}

Pulsars rotate with periodicities whose stability exceeds that of
any terrestrial clock. Pulsar frequencies and their time derivatives
are thereby by far the most accurately measured properties of compact
stars, whereas all other information about them is subject to much
greater uncertainties. Correspondingly it is extremely tempting to
exploit this information in order to learn about their internal structure
and in particular whether they consist of novel phases of dense matter
that might contain deconfined quarks \cite{Alford:2007xm}. This requires
the development of unique signatures that connect particular microscopic
properties to the macroscopic data. 

\begin{figure}
\flushright\includegraphics[scale=0.35]{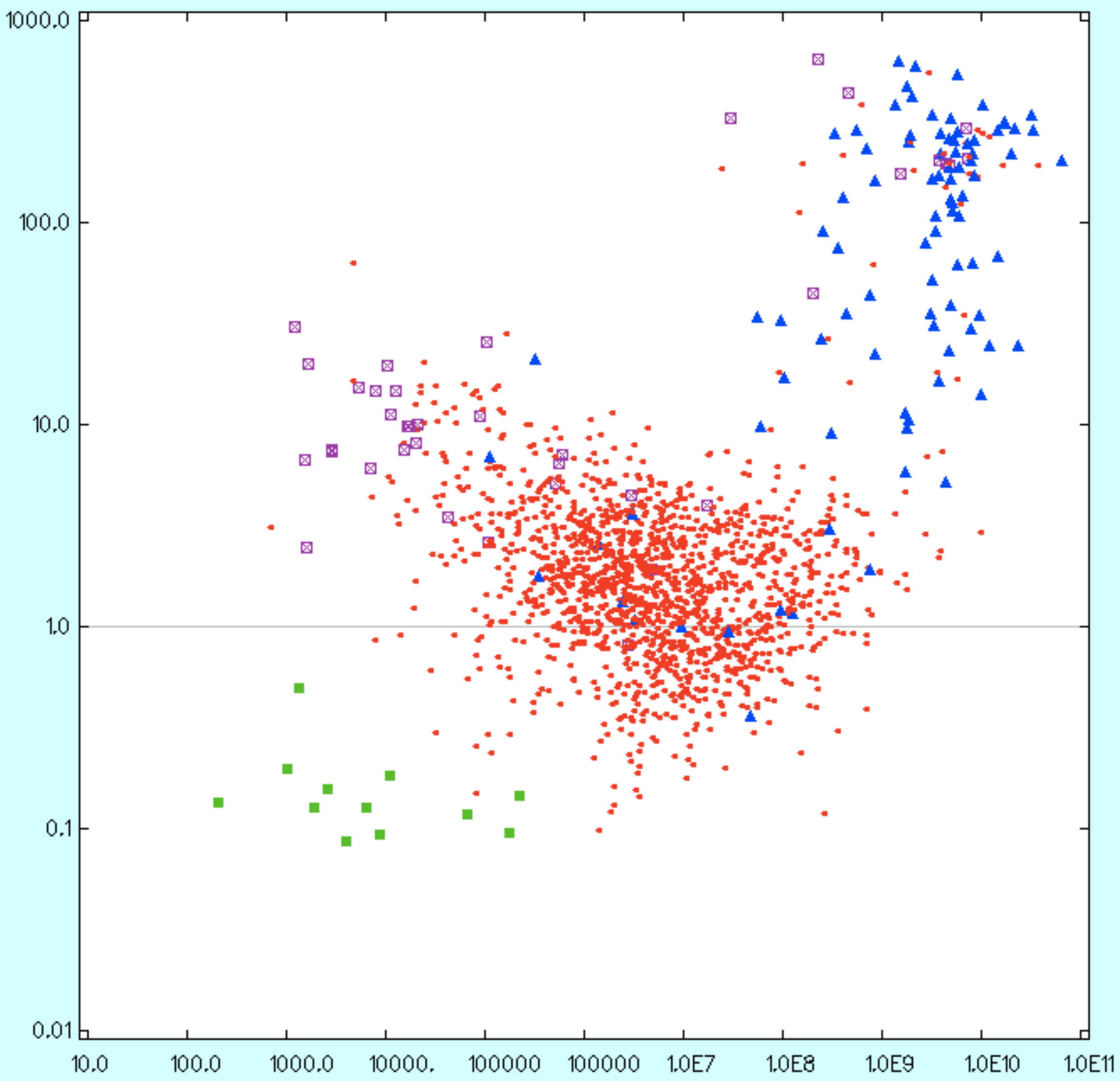}\vspace*{1mm}

\caption{\label{fig:pulsar-frequencies}Rotation frequencies of observed pulsars
versus their approximate spin-down age from the ATNF pulsar catalogue
\cite{Manchester:2004bp}.}

\flushleft \vspace*{-6.2cm} \hspace*{-0.1cm} $\nu\,[Hz]$ \\
\vspace*{3.85cm} \hspace*{4.1cm} $age\,[y]$ \\
\vspace*{1.3cm}
\end{figure}

Pulsar frequencies can change over time both by accretion of matter
that transfers angular momentum from a companion star and by the emission
of gravitational radiation due to oscillations of the star. A particularly
interesting class of oscillation modes are r-modes \cite{Papaloizou:1978zz,Lindblom:1999yk,Andersson:2000mf}
which are counter-rotating modes of a rotating star and are in the
absence of viscous damping unstable at all rotation frequencies \cite{Andersson:1997xt}.
This instability transforms rotational into gravitational wave energy
and leads to an exponential rise of the r-mode amplitude. When viscous
damping is taken into account the star is stable at low frequencies
but there remains an instability region at high frequencies \cite{Lindblom:1998wf,Jaikumar:2008kh}.
If this instability is stopped at a large amplitude, r-modes are a
strong and continuous source of gravitational waves and could provide
an extremely efficient mechanism for the spin-down of a young compact
star \cite{Owen:1998xg,Andersson:1998ze}. Observational data for
spin frequencies of pulsars, that spin down and allow the determination
of an approximate age associated to their spin-down rate%
\footnote{The spin-down age is determined by the assumption that magnetic dipole
breaking, which features a qualitatively similar behavior as gravitational
wave emission, dominates the spin-down. It can thereby only give a
rough order of magnitude estimate for the age. Age estimates are,
in particular, not available for stars that currently spin up and
which are correspondingly not included in the plot. However, for the
youngest stars there are independent age determinations from the observation
of the corresponding supernova remnant that qualitatively agree with
these estimates.%
}, is shown in fig.~\ref{fig:pulsar-frequencies}. Whereas observed
old pulsars in binary systems can spin nearly as fast as the maximum
Kepler frequency, above which the binding force cannot counteract
the centrifugal pseudo-force anymore, and can feature rotation periods
in the milli-second range, younger stars are far below this limit.
This is surprising since in their creation during a supernova a significant
fraction of the angular momentum of the initial star should be taken
over by the much smaller compact core which therefore should dramatically
spin up. This naive assumption is backed up by explicit analyses where
millisecond rotation frequencies at birth are indeed possible \cite{Ott:2005wh}.
If r-modes spin down compact stars on time scales shorter than the
age of the youngest observed pulsar, the lower boundary of the instability
window should give an upper limit for the maximum rotation frequency
of young compact stars%
\footnote{Note that the final frequency of the spin-down evolution can lie below
the minimum of the instability boundary since the damping of the r-mode
can take some time to complete, even after it enters the stable regime.%
}. 

R-modes are also relevant for the case of older stars in binaries
that are spun up by accretion since they generally limit the maximum
possible rotation frequency of a star to values substantially below
the Kepler frequency. A challenging finding is that, in contrast to
purely hadronic stars, more exotic possibilities like selfbound strange
stars \cite{Madsen:1999ci}, hybrid stars \cite{Jaikumar:2008kh}
or stars where hyperons are present in the core \cite{Lindblom:2001hd}
can feature so-called {}``stability windows'' where over a range
of intermediate temperatures the r-mode instability is absent up to
rather high frequencies. The observation of stars rotating at such
frequencies could therefore provide evidence for exotic phases in
their interior. In this context the masses and radii of stars provide
further important information. The recent precise measurement of a
heavy compact star with $M\approx2M_{\odot}$ \cite{Demorest:2010bx,Ozel:2010bz}
puts constraints on the presence of exotic phases since such phases
lead to a softening of the equation of state which in general leads
to a smaller maximum mass that is achievable for such an equation
of state. In combination with pulsar data this should lead to more
restrictive bounds on the possible presence of certain forms of matter
in compact stars.

A major problem for the extraction of information on the composition
of compact stars from observational data is the huge theoretical uncertainty
in the equation of state of dense matter and its transport properties.
This holds both for the hadronic side, where nuclear data is only
available at low densities and large proton fractions, and also for
hypothetical phases of quark matter, since QCD as the fundamental
theory of strong interactions cannot be solved so far in this non-perturbative
regime. Another big uncertainty factor is the crust of the neutron
star since although its microscopic physics is in general more constrained
by experimental data than the high density phases in the core, its
structural complexity limits so far a complete description. Simplified
estimates suggest that the crust could have a strong impact on the
damping of r-mode oscillations via surface rubbing at the crust-core
interface \cite{Bildsten:2000ApJ...529L..33B}. However, as in all
present analyses, r-modes are considered as solutions to the hydrodynamics
equations of an ideal fluid. When certain regions feature viscosities
that are enhanced by orders of magnitude, like at the crust-core interface,
the r-mode profile should strongly change in these regions and the
reduced amplitude there would result in a considerably weaker damping.
Therefore, we neglect crust effects in our present analysis, but a
more complete understanding of these effects in the future is desirable.
Unfortunately due to all this, even if two phases feature significant
qualitative differences these are often overshadowed by the huge quantitative
uncertainties in the detailed microscopic properties of either of
them. However, it was previously observed that certain features, like
the important case of the minimum of the instability region can be
surprisingly insensitive to quantitative details of the considered
models \cite{Lindblom:1998wf}. If such statements can be substantiated
this could allow us to devise robust signatures of the qualitative
features that can be stringently tested with present and forthcoming
astrophysical data.

To this end, we study in this paper the instability regions of small
amplitude r-modes in detail. In contrast to previous treatments that
studied particular star models numerically, we derive general analytic
results that are valid for stars consisting of various forms of matter
and compare these to numeric evaluations for realistic equations of
state. We find that although the form of the instability region can
be qualitatively distinct for the different forms of matter, many
aspects of the instability regions are extremely insensitive to the
detailed unknown microscopic input, like the transport coefficients
of a given phase. Moreover, we reveal the parametric dependence on
the underlying microscopic parameters where such dependences are significant.
We study explicitly the cases of neutron stars, strange stars as well
as hybrid stars and in view of the recent discovery of a $2\, M_{\odot}$
compact star, we generally apply only equations of state that can
accommodate such a heavy star. We find that the instability regions
of the hybrid stars are almost indistinguishable from those of neutron
stars if the size of the quark matter core is smaller than roughly
half of the star's radius. Further, we also study the case of an ultra-heavy
neutron star where direct Urca reactions are allowed in the core and
find that it features a notch at the right hand side of the instability
region. Finally, in addition to the dominant $m=2$ r-mode we also
consider higher multipoles and note that they could easily be excited
and become important for the evolution of the star.

In a companion paper \cite{Alford:2011pi} we show that due to the
strong increase of the bulk viscosity with amplitude \cite{Alford:2010gw,Madsen:1992sx,Reisenegger:2003pd,Bonacic:2003th}
the r-mode instability is only present at sufficiently small amplitudes
and the exponential r-mode growth is eventually saturated at finite
amplitudes so that r-modes could indeed provide a viable mechanism
for both the spin-down of young stars and the frequency limit of old,
accreting stars.

\section{Star models and r-modes}

\subsection{Static star models}

\begin{figure}
\includegraphics{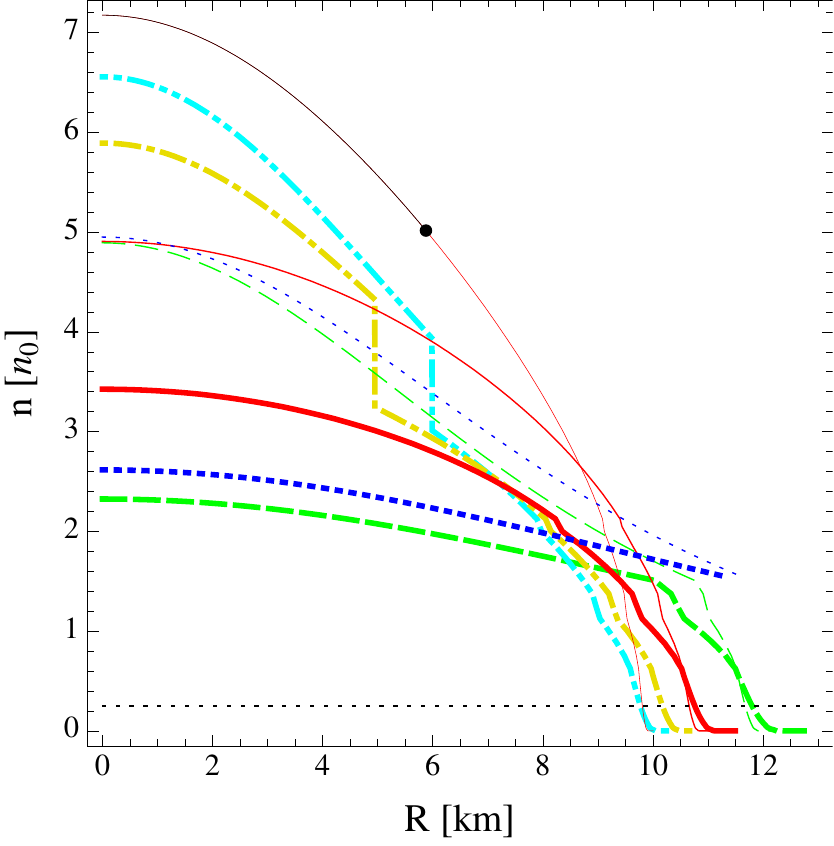}

\caption{\label{fig:density profiles}The density profiles of the star models
considered in this work. The solid lines represent neutron star models
with an APR equation of state, the dotted lines represent strange
stars with a bag model equation of state and the dashed, dot-dot-dashed
and dot-dashed lines represent hybrid star models with a large, medium
and small quark matter core respectively. Thick lines represent $1.4\, M_{\odot}$
stars and thin lines represent massive $2\, M_{\text{\ensuremath{\odot}}}$
stars. In contrast to strange stars that are basically homogeneous,
stars that contain hadronic matter have a very strong density dependence
that extends over 14 orders of magnitude reflected by the near zero
segments in this plot. The very thin solid curve presents the maximum
neutron star model for the APR equation of state $\sim2.2\, M_{\odot}$
where hadronic direct Urca processes are allowed to the left of the
dot. The dotted horizontal line denotes the density $n=n_{0}/4$ chosen
as the beginning of the crust whose contribution is not taken into
account in the damping time integrals below.}

\end{figure}

The analysis of compact star oscillations and their damping requires
as a first step the stable equilibrium configuration of the star.
In this section we will discuss the considered star models that are
used later on. The equilibrium star configuration is determined by
the gravitational equations for a fluid sphere. We employ here the
general relativistic Tolman-Oppenheimer-Volkov (TOV) equations \cite{Tolman:1939jz}.
The latter require the equation of state of neutral and $\beta$-equilibrated
dense matter. The recent measurement of a neutron star with the large
mass $1.97\pm0.04\, M_{\odot}$ \cite{Demorest:2010bx,Ozel:2010bz}
puts bounds on the equation of state of dense matter. We generally
study equations of state that can accommodate such a heavy star. We
consider three qualitatively different classes and study in each case
both a star model with a standard value of $1.4\, M_{\odot}$ and
one with a large mass $2\, M_{\odot}$:
\begin{enumerate}
\item Neutron stars (NS) are obtained for an equation of state that is hadronic
at all relevant densities. Whereas the above maximum mass does not
pose problems for neutron stars obtained from most hadronic equations
of state, it seem nearly impossible to obtain such heavy stars which
contain a significant amount of hyperonic matter \cite{Ozel:2010bz}.
Therefore we do not study this possibility here and consider only
stars consisting of neutrons, protons and electrons as well as muons
at sufficiently high density. We also neglect the possible presence
of hadronic pairing which in general significantly reduces the bulk
viscosity \cite{Haensel:2000vz,Haensel:2001mw}. Such pairing is only
realized in certain shells within the star and as long as they are
not large this might not qualitatively change the results obtained
here. However, the intricate dynamics of a two-component fluid can
change this simplistic picture \cite{Haskell:2009fz}. For our numerical
results we employ the equilibrated equation of state by Akmal, Pandharipande
and Ravenhall (APR) \cite{Akmal:1998cf} which relies on a potential
model that reproduces scattering data in vacuum supplemented by a
model for three-body interactions in order to reproduce the saturation
properties at nuclear densities. As a low density extension of the
APR data we use \cite{Baym:1971pw,Negele:1971vb}. Furthermore, we
study in this class also a neutron star with an ultra-high mass $2.21\, M_{\odot}$
close to the mass limit for the APR equation of state, since in this
case direct Urca interactions are possible in the interior of the
star leading to a significantly enhanced weak rate, cf. e.g. \cite{Alford:2010gw}.
\item Hybrid stars (HS) with an outer hadronic part and a core of quark
matter are obtained from an equation of state where at some density
the effective degrees of freedom change from hadrons to quarks. In
general the equation of state of interacting quark matter is unknown
and there are only hints from the perturbative regime \cite{Fraga:2001id,Fraga:2004gz,Kurkela:2009gj}
or model studies. We use the simple quartic parameterization for the
equation of state of ungapped 3-flavor quark matter \cite{Alford:2004pf,Alford:2010gw}
in terms of the individual quark chemical potentials $\mu_{d}$, $\mu_{u}$
and $\mu_{s}$ and the electron chemical potential $\mu_{e}$ \begin{align}
p_{par} & =\frac{1-c}{4\pi^{2}}\left(\mu_{d}^{4}+\mu_{u}^{4}+\mu_{s}^{4}\right)-\frac{3m_{s}^{2}\mu_{s}^{2}}{4\pi^{2}}\label{eq:quark-eos-model}\\
 & \quad+\frac{3m_{s}^{4}}{32\pi^{2}}\left(3+4\log\!\left(\frac{2\mu_{s}}{m_{s}}\right)\right)-{\cal B}+\frac{\mu_{e}^{4}}{12\pi^{2}}\nonumber \end{align}
where $c$, $m_{s}$ and $B$ are effective model parameters that
incorporate some effects of the strong interactions between the quarks.
From this equation of state we find the $\beta$-equilibrated and
charge neutral ground state which depends on a single quark number
chemical potential $\mu_{q}$. We use the general form since the computation
of transport properties below requires susceptibilities around the
equilibrium state. Within the parameterization eq. (\ref{eq:quark-eos-model})
the recent measurement of a heavy star strongly restricts the equation
of state so that only equations of state that are strongly interacting
($c>0.3$) are compatible \cite{Alford:2004pf,Ozel:2010bz}. These
are equations of state where the transition to quark matter is at
rather low values of the baryon number $\lesssim1.5\, n_{0}$, where
$n_{0}$ is nuclear matter density, and we choose here one where it
occurs at $1.5\, n_{0}$. Since the APR equation of state happens
to be very similar to the above form eq. (\ref{eq:quark-eos-model})
so that even multiple transitions are possible \cite{Alford:2004pf},
we do not expect the transition density to be a robust result that
is independent of the considered equations of state. Therefore we
study here in addition also two $1.4\, M_{\odot}$ hybrid star models
obtained with quark equations of state that cannot accommodate a heavy
star when combined with the APR equation of state. Ona has a small
quark core, obtained for a transition density of $3.25\, n_{0}$,
and the other a medium-sized quark core, obtained for a transition
density of $3\, n_{0}$.\\
There are many possible phases of quark matter that feature various
color superconducting pairing patterns. Here we do not explicitly
study star models with pairing, but note that the parameterization
eq. (\ref{eq:quark-eos-model}) can also describe superconducting
matter. For our hybrid star models we make the assumption of local
charge neutrality, excluding the possibility of a mixed phase and
thereby circumventing the description of the wealth of possible geometric
structures of such a mixture.
\item Strange stars (SS) that are self-bound could exist according to the
strange matter hypothesis \cite{Witten:1984rs} that the true ground
state of strongly interacting matter is 3-flavor quark matter. For
strange stars we use the same equation of state eq. (\ref{eq:quark-eos-model}),
but use parameter sets that realize the strange matter hypothesis.
In contrast to the case of hybrid stars large mass strange stars are
possible even for $c=0$ and due to our ignorance of the precise form
of the interacting equation of state we choose this value to keep
our model as simple as possible. For a strange quark mass of $m_{s}=150$
MeV stable strange stars exist in this case for bag constants $B_{lim}<\left(158\, MeV\right)^{4}$,
whereas heavy strange stars exist for this mass value only for bag
constants $B\lesssim\left(140\, MeV\right)^{4}$. Lower effective
quark mass values or changes in the quartic term ($c\neq0$) relax
these bounds. 
\end{enumerate}
We note already at this point that although the detailed parameters
we chose here for our star models are rather arbitrary, we will give
general analytic expressions below that will reveal the dependence
of our main results on the various model parameters. The characteristic
parameters of the considered star models are given in table \ref{tab:star-models}.
As is a well known property, the radii of the different models vary
only moderately with the mass for masses of $1$ to $2\, M_{\odot}$.
The density profiles are shown in fig.~\ref{fig:density profiles}.
The core densities of these star models range from below $3\, n_{0}$
to more than $7\, n_{0}$. In contrast to neutron and hybrid stars
that vary in density by 14 orders of magnitude, strange stars feature
a roughly constant density profile.

\begin{table}[H]
\begin{tabular}{|c|c|c|c|c|c|c|}
\hline 
 & $M\left[M_{\odot}\right]$ & $M_{core}\left[M_{\odot}\right]$ & $R\left[km\right]$ & $n_{c}\left[n_{0}\right]$ & $\left\langle n\right\rangle \left[n_{0}\right]$ & $\Omega_{K}\left[kHz\right]$\tabularnewline
\hline 
NS & $1.4$ & $\left(1.39\right)$ & $11.5$ & $3.43$ & $1.58$ & $6.02$\tabularnewline
\hline 
 & $2.0$ & $\left(1.99\right)$ & $11.0$ & $4.91$ & $2.46$ & $7.68$\tabularnewline
\hline 
 & $2.21$ & $0.85$ & $10.0$ & $7.17$ & $3.37$ & $9.31$\tabularnewline
\hline 
SS & $1.4$ & $-$ & $11.3$ & $2.62$ & $1.91$ & $6.17$\tabularnewline
\hline 
 & $2.0$ & $-$ & $11.6$ & $4.95$ & $2.43$ & $7.09$\tabularnewline
\hline 
HS & $1.4$ $\left(S\right)$ & $0.38^{*}$ & $10.8$ & $5.89$ & $1.85$ & $6.61$\tabularnewline
\hline 
 & $1.4$$\left(M\right)$ & $0.66^{*}$ & $10.3$ & $6.66$ & $2.09$ & $7.06$\tabularnewline
\hline 
 & $1.4$$\left(L\right)$ & $1.06$ & $12.7$ & $2.32$ & $1.17$ & $5.16$\tabularnewline
\hline 
 & $2.0$ & $1.81$ & $12.2$ & $4.89$ & $1.84$ & $6.62$\tabularnewline
\hline
\end{tabular}

\caption{\label{tab:star-models}Results of the considered models of neutron
stars (NS), strange stars (SS) and hybrid stars (HS). Shown are the
mass of the star $M$, the mass of the core $M_{core}$, the radius
$R$, the baryon density at the center of the star $n_{c}$ given
in units of nuclear saturation density $n_{0}$, the average density
$\left\langle n\right\rangle $ and the Kepler frequency $\Omega_{K}$.
The neutron stars were obtained by solving the relativistic TOV equations
for catalyzed neutron matter using the APR equation of state \cite{Akmal:1998cf}
with low density extension \cite{Baym:1971pw,Negele:1971vb} and the
strange stars with a quark gas bag model with $c=0$, $m_{s}=150\, MeV$
and a bag parameter $B=\left(138\, MeV\right)^{4}$. Large mass hybrid
stars are only found when strong interaction corrections are considered,
cf. \cite{Alford:2004pf}, and we find a $2\, M_{\odot}$ star for
$c=0.4$, $m_{s}=140\, MeV$, $B=\left(137\, MeV\right)^{4}$. The
additional two $1.4\, M_{\odot}$ hybrid models with smaller cores,
marked with an asterisk, result from equations of state that do not
allow large mass models. They correspond to $c=0$, $m_{s}=150\, MeV$,
$B=\left(164.5\, MeV\right)^{4}$ and $\left(171.5\, MeV\right)^{4}$
which are chosen to obtain transition densities of $3n_{0}$ and $3.25n_{0}$,
respectively.}

\end{table}

\subsection{R-mode profile}

The analysis of oscillation modes of compact stars requires the solution
of the corresponding hydrodynamics equations \cite{Andersson:2000mf}.
Whereas the solution of the general relativistic equations for static
star models is straightforward, the corresponding dynamic equations
are more involved and explicit analytic expressions for the r-mode
oscillation are only available in non-relativistic approximation.
Yet numerical analyses of the general relativistic equations show
that except for very compact stars the relativistic corrections are
moderate \cite{Lockitch:2000aa,Lockitch:2002sy,Ruoff:2001wr}. In
the non-relativistic case the current state of the art is the comprehensive
analysis \cite{Lindblom:1999yk}. It might therefore seem more consistent
to employ Newtonian equations for the static star models as well,
but we prefer to perform the necessary approximation for the oscillation
at least around the correct equilibrium configuration. In particular
since we study heavy quark and hybrid stars where the different approximations
could differ. In particular in the large amplitude regime, a consistent
general relativistic analysis would clearly be desirable. Furthermore,
so far these analyses assume that the oscillation modes are solutions
of an ideal fluid. For these modes the damping is then computed in
a second step.

R-modes are normal oscillations of rotating stars and correspondingly
they require as a first step the solution of a uniformly rotating
stellar model. Since neutron stars are cold, dense systems we assume
a barotropic fluid where the pressure $p$ is only a function of the
energy density $\rho$. The hydrodynamic Euler equation for this spherical
system, determining the enthalpy $h$ and the equation for the gravitational
potential $\Phi$ of the star, have to be solved with appropriate
boundary conditions \cite{Lindblom:1999yk}. To simplify the demanding
analysis a slow rotation expansion is performed and, since the density
fluctuation of the r-mode vanishes to leading order in the expansion,
the computation of bulk viscosity damping times requires an expansion
of $h$, $\Phi$ and of the energy density $\rho$ to next-to-leading
order

\[
X\!\left(r,\cos\theta\right)=X_{0}\!\left(r\right)+X_{2}\!\left(r,\cos\theta\right)\left(\frac{\Omega^{2}}{\pi G\bar{\rho}_{0}}\right)+\cdots\]
where $X$ stands for either of the three quantities, $\Omega$ is
the angular velocity of the rotation and $\bar{\rho}_{0}$ the average
energy density of the corresponding non-rotating star. The main effect
of the rotational corrections is a flattening of the star due to centrifugal
forces.

The next step is the search for eigenmodes of the rotating star in
a linear low amplitude approximation, e.g. for the conserved baryon
number $\Delta n\ll\bar{n}$. They can completely be described by
the change in the gravitational potential $\delta\Phi$ and the hydrodynamical
perturbation

\[
\delta U=\frac{\delta p}{\rho}-\delta\Phi\:.\]
These are likewise expanded to next-to-leading order in $\Omega$
in the form

\[
\delta X\!\left(r,\cos\theta\right)=R^{2}\Omega^{2}\left(\delta X_{0}\!\left(r\right)\!+\!\delta X_{2}\!\left(r,\cos\theta\right)\frac{\Omega^{2}}{\pi G\bar{\rho}_{0}}\!+\!\cdots\right)\]
where $\delta X$ stands again either for $\delta\Phi$ or $\delta U$
and $R$ is the radius of the static star. The potentials obey complicated
differential equations with corresponding boundary conditions given
in \cite{Lindblom:1999yk}. The boundary conditions require that the
oscillation frequencies $\omega_{r}$ in the rotating and $\omega_{i}$
in the inertial frame are connected to the rotation frequency $\Omega$
via 

\begin{equation}
\omega_{r}\equiv\omega=\kappa(\Omega)\Omega\quad,\quad\omega_{i}=\omega_{r}-m\Omega\label{eq:frequency-connection}\end{equation}
where the parameter $\kappa$ can likewise be expanded in $\Omega$

\[
\kappa=\kappa_{0}+\kappa_{2}\frac{\Omega^{2}}{\pi G\bar{\rho}_{0}}+\cdots\]
We study classical r-modes which are a one parameter class of eigenmode
solutions, $l=m$, that are to leading order determined by $\kappa_{0}=2/\left(m+1\right)$
and given by

\[
\delta U_{0}\!\left(\vec{r}\right)=\sqrt{\frac{m}{\pi\left(m\!+\!1\right)^{3}\left(2m\!+\!1\right)!}}\alpha\left(\frac{r}{R}\right)^{m+1}P_{m+1}^{m}\!\left(\cos\theta\right)\mathrm{e}^{im\varphi}\]
in terms of associated Legendre polynomials $P_{m+1}^{m}$ and with
the definition of the dimensionless amplitude $\alpha$ introduced
in \cite{Lindblom:1998wf}. For the lowest r-mode that couples to
gravitational waves with $m=2$ the oscillation frequency in the inertial
frame is given by $\omega_{i}=-4/3\,\Omega$, corresponding to a counter-rotating
flow. The leading order gravitational potential obeys a differential
equation that is given in appendix \ref{sec:Analytic-r-mode} where
its analytic solution is given in the special case that the star is
of uniform density. In general it requires a numeric solution and
then completely determines the r-mode to this order.

\begin{figure}
\includegraphics{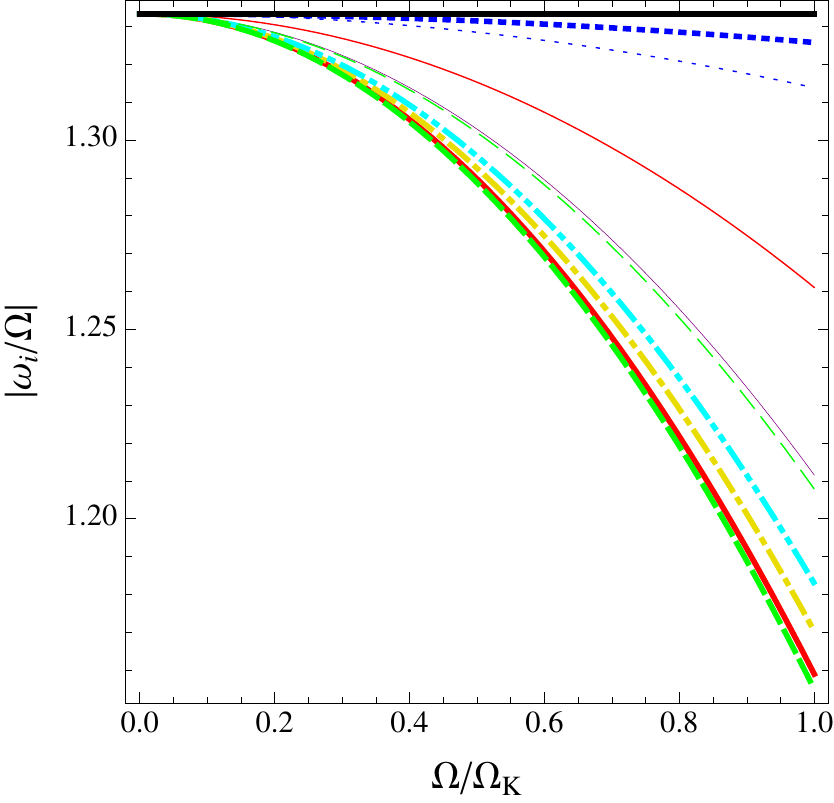}

\caption{\label{fig:oscillation frequency}Connection of the oscillation frequency
$\omega_{i}$ of the r-mode in the inertial frame to the rotation
frequency of the considered star models to next to leading order in
the $\Omega$-expansion. The horizontal line shows the leading order
result and the conventions for the other curves are the same as in
fig.~\ref{fig:density profiles}.}

\end{figure}

At next-to-leading order there are two qualitatively different effects:
First the connection between oscillation and rotation frequency as
described by the parameter $\kappa$ becomes non-linear via a non-vanishing
value of $\kappa_{2}$ that has to be obtained numerically from a
corresponding differential equation \cite{Lindblom:1999yk}. In fig.~\ref{fig:oscillation frequency}
the solution is shown for the star models discussed here. Whereas
the corrections are small for quark stars, at large frequency they
can become sizable for hadronic and hybrid stars. The second effect
is the change of the potentials arising as solutions of rather involved
partial differential equations. At next-to-leading order they feature
nontrivial radial and angular dependences that are not described by
simple power laws or spherical harmonics anymore.

We are interested in the amplification of these modes due to gravitational
waves and their viscous damping which are described by the energy
dissipation \begin{align}
\frac{dE}{dt}= & -\omega\left(\omega-m\Omega\right)^{2m+1}\left|\delta J_{mm}\right|^{2}\nonumber \\
 & -\int d^{3}x\left(2\eta\delta\sigma^{ab}\delta\sigma_{ab}+\zeta\delta\sigma\delta\sigma^{*}\right)\label{eq:dissipated-energy}\end{align}
where $\eta$ and $\zeta$ are the shear and bulk viscosity, respectively.
The fluctuations $\delta J_{mm}$ and $\delta\sigma_{ab}$ couple
to gravitational waves and shear viscosity, respectively. The fluctuation
$\delta\sigma\equiv\vec{\nabla}\cdot\delta\vec{v}$ of an r-mode oscillation
which is subject to dissipation via bulk viscosity reads

\begin{equation}
\delta\sigma=-i\frac{2AR^{2}\Omega^{3}}{m+1}\left(\delta U_{0}+\delta\Phi_{0}+\cdots\right)+O\!\left(\Omega^{5}\right)\label{eq:delta-sigma}\end{equation}
where $A$ denotes the inverse speed of sound

\begin{equation}
A\equiv\left.\frac{\partial\rho}{\partial p}\right|_{0}\label{eq:A-parameter}\end{equation}
evaluated at equilibrium and the dots represent several further terms
of next-to-leading order in $\Omega$ given in \cite{Lindblom:1999yk}.
The terms in the parenthesis depend on the expansion coefficients
of the potentials and on the connection parameter $\kappa$, but are
by virtue of the expansion independent of frequency. The fluctuation
$\delta\sigma$ is finally connected to the fluctuations of energy
$\Delta\rho$ and baryon number density $\Delta n$ via their continuity
equations 

\[
\left|\delta\sigma\right|=\left|\vec{\nabla}\cdot\delta\vec{v}\right|=\kappa\Omega\left|\frac{\Delta\rho}{\bar{\rho}}\right|=\kappa\Omega\left|\frac{\Delta n}{\bar{n}}\right|\]
Following \cite{Jaikumar:2008kh}, in our numerical analysis we will
consider the change of $\kappa$ at second order, but because of the
involved numerics \cite{Lindblom:1999yk}, we will not take into account
all explicit second order terms in eq. (\ref{eq:delta-sigma}). A
standard approximation \cite{Lindblom:1998wf} is to replace the Lagrangian
density fluctuation by the Eulerian one which corresponds to neglecting
the terms denoted by the parenthesis in eq. (\ref{eq:delta-sigma}).
E.g. the density fluctuation for a $m=2$ r-mode reads then to leading
order

\begin{equation}
\left|\frac{\Delta\rho}{\bar{\rho}}\right|\approx\left|\frac{\delta\rho}{\bar{\rho}}\right|=\sqrt{\frac{8}{189}}\alpha AR^{2}\Omega^{2}\left(\left(\frac{r}{R}\right)^{3}\!+\!\delta\Phi_{0}\left(r\right)\right)Y_{3}^{2}\!\left(\theta,\phi\right)\label{eq:r-profile}\end{equation}
in terms of the spherical harmonics $Y_{3}^{2}$. However, we will
give general semi-analytic expressions below that \emph{are} valid
to full next to leading order. These show that the influence of the
neglected terms on important aspects of the instability regions is
rather mild.

\section{Viscosities of dense matter}

\subsection{General expressions}

The bulk viscosity describes the local dissipation of energy in a
fluid element which arises from the global oscillation mode of the
star within a compression and rarefaction cycle. The integration over
the whole star yields then the corresponding energy dissipation of
the mode as will be discussed in the next section. Recently the bulk
viscosity of large amplitude oscillations \cite{Madsen:1992sx} has
been studied in detail \cite{Alford:2010gw}. There it was shown that
large amplitude oscillations are in general considerably more strongly
damped and this mechanism can thereby saturate unstable r-modes at
finite amplitudes as will be discuss in a companion article. In this
article, however, we restrict ourselves to the subthermal regime $\mu_{\Delta}\ll T$
which determines whether small amplitude r-modes are initially unstable. 

The bulk viscosity is maximal when the external oscillation frequency
matches the time scale of the microscopic interactions that establish
equilibrium. The relevant interactions in the case of star oscillations
are slow weak processes. The parametric form of the beta-equilibration
rate is given by

\begin{equation}
\Gamma^{(\leftrightarrow)}=-\tilde{\Gamma}T^{\delta}\mu_{\Delta}\left(1+\sum_{j=1}^{N}\chi_{j}\left(\frac{\mu_{\Delta}^{2}}{T^{2}}\right)^{j}\right)\label{eq:gamma-parametrization}\end{equation}
where $\mu_{\Delta}\equiv\sum_{i}\mu_{i}-\sum_{f}\mu_{f}$ is the
quantity that is driven out of equilibrium due to the oscillations
and its re-equilibration leads to the bulk viscosity. In the subthermal
regime, $\mu_{\Delta}\ll T$, the non-linear terms in the rate of
the equilibration processes can be neglected, leading to the general
analytic result for the subthermal bulk viscosity \cite{Alford:2010gw}

\begin{equation}
\zeta^{<}=\frac{C^{2}\tilde{\Gamma}T^{\delta}}{\omega^{2}+\left(B\tilde{\Gamma}T^{\delta}\right)^{2}}=\zeta_{max}^{<}\frac{f}{1+f^{2}}\label{eq:sub-viscosity}\end{equation}
in terms of the reduced weak rate $\tilde{\Gamma}$ and the strong
susceptibilities $B$ and $C$

\begin{equation}
C\equiv\bar{n}\left.\frac{\partial\mu_{\Delta}}{\partial n}\right|_{x}\quad,\quad B\equiv\frac{1}{\bar{n}}\left.\frac{\partial\mu_{\Delta}}{\partial x}\right|_{n}\label{eq:susceptibilities}\end{equation}
as well as the oscillation frequency $\omega$ (in the rotating frame)
and temperature $T$. It has a characteristic resonant form and as
long as the combination of susceptibilities $C^{2}/B$ does not vary
too quickly with temperature, the sub-thermal viscosity has a maximum
\begin{equation}
\zeta_{max}^{<}=\frac{C^{2}}{2\omega B}\quad\mathrm{at}\quad T_{max}=\left(\frac{\omega}{\tilde{\Gamma}B}\right)^{\frac{1}{\delta}}\label{eq:max-viscosity}\end{equation}
and simple asymptotic limits 

\[
\zeta^{<}=\zeta_{max}^{<}\cdot\left\{ \begin{array}{cc}
\!\! f & \quad\left(f\ll1\right)\\
\!\!\frac{1}{f} & \quad\left(f\gg1\right)\end{array}\right.\]
where $f\equiv B\tilde{\Gamma}T^{\delta}/\omega$ can be identified
as the feedback term in the differential equation that determines
the damped oscillation \cite{Alford:2010gw}.

The shear viscosity arises from strong or electromagnetic interactions.
In contrast to the bulk viscosity, the shear viscosity of dense matter
is independent of the frequency of an external oscillation and approximately
depends on temperature via a simple power law. Shear viscosity becomes
large at low temperatures and therefore it is the dominant process
for damping of the r-modes of cooler stars. Thereby, to leading order
it can be parameterized as

\begin{equation}
\eta=\tilde{\eta}T^{-\sigma}\label{eq:shear-parametrization}\end{equation}
by simply factoring out the temperature dependence with exponent $\sigma$.
In general several processes can contribute so that the full shear
viscosity can approximately be written as a sum of such power laws
for the individual processes.

\subsection{Viscosities for the considered forms of matter}

In the following we discuss the viscosities for the phases of dense
matter presented in the considered classes of compact stars. We note
however, that the general results given here likewise apply to more
complicated forms of matter like hyperonic and/or superfluid nuclear
matter as well as various forms of superconducting quark matter.

\subsubsection{Bulk viscosity}

The general expression for the bulk viscosity eq. (\ref{eq:sub-viscosity})
depends on the weak rates. In the case of hadronic matter in the absence
of hyperons the weak equilibration occurs via the Urca channel

\[
n\to p+e^{-}+\bar{\nu}_{e}\quad,\quad p+e^{-}\to n+\nu_{e}\]
There are two qualitatively different cases depending on whether the
direct process \cite{Haensel:1992zz,Reisenegger:1994be} is possible
or only the modified version \cite{Friman:1978zq,Sawyer:1989dp,Reisenegger:1994be}
where by-stander nucleons are necessary to satisfy energy momentum
conservation. The latter represents a particular strong interaction
vertex correction to the above process. However, from the point of
view of the weak interaction these different processes belong to the
same channel. Further, there is in principle also a corresponding
process with muon instead of electrons. Since the phase space and
thereby the rate of this process is strongly suppressed compared to
the electron version at moderate densities we do not consider it here. 

In strange quark matter the dominant channel for beta equilibration
is the non-leptonic flavor changing process

\[
s+u\leftrightarrow d+u\]
whereas the corresponding quark Urca processes are parametrically
suppressed in $T/\mu\ll1$. The weak parameters $\tilde{\Gamma}$,
$\delta$ and $\chi$ in the parameterization of the weak rate eq.
(\ref{eq:gamma-parametrization}) are given on the left panel of table
\ref{tab:weak-parameters}. 

\begin{table*}
\begin{minipage}[t]{0.6\textwidth}%
\begin{tabular}{|c|c|c|c|c|c|}
\hline 
Weak process & $\tilde{\Gamma}\,\bigl[\mathrm{MeV}^{(3-\delta)}\bigr]$ & $\delta$ & $\chi_{1}$ & $\chi_{2}$ & $\chi_{3}$\tabularnewline
\hline 
quark non-leptonic & $6.59\!\times\!10^{-12}\,\Bigl(\frac{\mu_{q}}{300\,\mathrm{MeV}}\Bigr)^{5}$ & $2$ & $\frac{1}{4\pi^{2}}$ & $0$ & $0$\tabularnewline
\hline 
hadronic direct Urca & $5.24\!\times\!10^{-15}\left(\!\frac{x\, n}{n_{0}}\!\right)^{\!\frac{1}{3}}$ & $4$ & $\frac{10}{17\pi^{2}}$ & $\frac{1}{17\pi^{4}}$ & $0$\tabularnewline
\hline 
hadronic modified Urca & $4.68\!\times\!10^{-19}\left(\!\frac{x\, n}{n_{0}}\!\right)^{\!\frac{1}{3}}$ & $6$ & $\frac{189}{367\pi^{2}}$ & $\frac{21}{367\pi^{4}}$ & $\frac{3}{1835\pi^{6}}$\tabularnewline
\hline
\end{tabular}%
\end{minipage}%
\begin{minipage}[t]{0.4\textwidth}%
\begin{tabular}{|c|c|c|}
\hline 
Strong/EM process & $\tilde{\eta}\,\bigl[\mathrm{MeV}^{(3+\sigma)}\bigr]$ & $\sigma$\tabularnewline
\hline 
quark scattering & $1.98\!\times\!10^{9}\alpha_{s}^{-\frac{5}{3}}\left(\frac{\mu_{q}}{300\,\mathrm{MeV}}\right)^{\frac{14}{3}}$ & $\frac{5}{3}$\tabularnewline
\hline 
leptonic scattering & $1.40\!\times\!10^{12}\left(\frac{x\, n}{n_{0}}\right)^{\frac{14}{9}}$ & $\frac{5}{3}$\tabularnewline
\hline 
nn-scattering & $5.46\!\times\!10^{9}\left(\frac{\rho}{m_{N}n_{0}}\right)^{\frac{9}{4}}$ & $2$\tabularnewline
\hline
\end{tabular}%
\end{minipage}

\caption{\label{tab:weak-parameters}\emph{Left panel:} Parameters of the general
parameterization of the weak rate eq. (\ref{eq:gamma-parametrization})
for different processes of particular forms of matter which determine
the damping due to bulk viscosity. The coefficients $\chi_{i}$ parameterize
the non-linear dependence on the chemical potential fluctuation $\mu_{\Delta}$
arising in the suprathermal regime of the viscosity which is relevant
for large amplitude r-modes studied in \cite{Alford:2011pi} but which
are not relevant in this work. \emph{Right panel:} Parameters arising
in the parameterization eq. (\ref{eq:shear-parametrization}) of the
shear viscosity for different strong and electromagnetic interaction
processes. The leptonic and quark scattering arises from a non-Fermi
liquid enhancement due to unscreened magnetic interactions.}

\end{table*}
Further, the inverse speed of sound $A$, eq. (\ref{eq:A-parameter}),
as well as the strong susceptibilities $B$ and $C$, eq. (\ref{eq:susceptibilities}),
that parameterize the deviation from chemical equilibrium are required.
In order to make it easy to apply our general results to different
forms of interacting hadronic matter, we implement the APR equation
of state using the simple parameterization employed in \cite{Lattimer:1991ib}
to approximate the dependence of the energy per particle on the proton
fraction $x$ by a quadratic form \[
E\!\left(n,x\right)=E_{s}\!\left(n\right)+S\!\left(n\right)\left(1-2x\right)^{2}\]
where $E_{s}$ and $S$ are the corresponding energy for symmetric
matter and the symmetry energy for which we employ simple quadratic
fits to the APR data. For comparison we also consider the susceptibilities
in the approximation of a hadron gas which had been used in previous
analyses of the bulk viscosity \cite{Sawyer:1989dp,Haensel:1992zz,Jaikumar:2008kh}.
Likewise we compute these quantities for the generic, phenomenological
form of the equation of state of interacting quark matter eq. (\ref{eq:quark-eos-model}).
The resulting strong interaction parameters describing the response
of the different models are given in table \ref{tab:strong-parameters}. 

\begin{table*}
\begin{tabular}{|c|c|c|c|}
\hline 
 & $A$ & $B$ & $C$\tabularnewline
\hline 
hadronic matter & $m_{N}\left(\frac{\partial p}{\partial n}\right)^{-1}$ & $\frac{8S}{n}\!+\negthinspace\frac{\pi^{2}}{\left(4\left(1\!-\!2x\right)S\right)^{2}}$ & $4\!\left(1\!-\!2x\right)\!\left(n\frac{\partial S}{\partial n}\!-\!\frac{S}{3}\right)$\tabularnewline
\hline 
hadronic gas & $\frac{3m_{N}^{2}}{\left(3\pi^{2}n\right)^{\frac{2}{3}}}$ & $\frac{4m_{N}^{2}}{3\left(3\pi^{2}\right)^{\frac{1}{3}}n^{\frac{4}{3}}}$ & $\frac{\left(3\pi^{2}n\right)^{\frac{2}{3}}}{6m}$\tabularnewline
\hline 
quark matter (gas: $c=0$) & $3\!+\!\frac{m_{s}^{2}}{\left(1\!-\! c\right)\mu_{q}^{2}}$ & $\frac{2\pi^{2}}{3(1-c)\mu_{q}^{2}}\left(1\!+\!\frac{m_{s}^{2}}{12(1-c)\mu_{q}^{2}}\right)$ & $-\frac{m_{s}^{2}}{3(1-c)\mu_{q}}$\tabularnewline
\hline
\end{tabular}

\caption{\label{tab:strong-parameters}Strong interaction parameters, defined
in eqs. (\ref{eq:A-parameter}) and (\ref{eq:susceptibilities}),
describing the response of the particular form of matter. In the case
of interacting hadronic matter a quadratic ansatz in the proton fraction
$x$ parameterized by the symmetry energy $S$ is employed. The expressions
for a hadron and quark gas are given to leading order in $n/m_{N}^{3}$
respectively next to leading order in $m_{s}/\mu$.}

\end{table*}

\subsubsection{Shear viscosity}

In previous r-mode analyses the shear viscosity in hadronic matter
has been approximated by the contribution from strong hadron-hadron-scattering
using the fit in \cite{Cutler:1990} to the standard low density ($\lesssim n_{0}$)
data given in \cite{Flowers:1976ApJ...206..218F}

\begin{equation}
\eta_{n}=347\rho^{\frac{9}{4}}T^{-2}\frac{\mathrm{g}}{\mathrm{cm\, s}}\label{eq:hadronic-shear}\end{equation}
where $T$ is in units of Kelvin and $\rho$ is given in $\mathrm{g}/\mathrm{cm}^{3}$
. Extrapolating this fit to high densities relevant for neutron stars
overestimates the viscosity. The new evaluation in \cite{Shternin:2008es}
shows instead that due to a non-Fermi liquid enhancement arising from
the exchange of Landau-damped transverse photons, the main contribution
to the shear viscosity of hadronic matter, at temperatures relevant
to the spin-down evolution of the compact stars, comes from electron
scattering and is given by \begin{equation}
\eta_{e}=4.26\times10^{-26}(x\, n)^{\frac{14}{9}}T^{-\frac{5}{3}}\frac{\mathrm{g}}{\mathrm{cm\, s}}\label{eq:electron-shear}\end{equation}
where $T$ is in Kelvin and the baryon number density $n$ is in units
of $\mathrm{cm}^{-3}$. In the calculation of $\eta_{e}$ we have
only considered electron-electron and electron-proton scattering and
neglected the small effect of the muons to the shear viscosity. For
densities larger than nuclear matter saturation density this electron
contribution dominates over the hadronic one down to temperatures
below $10^{7}$ K. This region contains the part of the instability
region that is relevant for the spin-down evolution of stars. Therefore
in our main analysis we will completely neglect the hadronic component
of the shear viscosity. However, we will compare with the previous
form of the shear viscosity eq. (\ref{eq:hadronic-shear}) in order
to discuss the effect of our improved analysis on the instability
region%
\footnote{The leftmost part of the instability region below $\sim10^{7}$ K
only features an instability at very large frequencies. This part
would only be relevant for old stars that are spun up by accretion,
yet frequencies very close to the Kepler frequency are not reached
via this mechanism anyway, cf. fig.~\ref{fig:pulsar-frequencies},
due to turbulent effects.%
}. 

In the case of ungapped quark matter, the shear viscosity is dominated
by quark-quark scattering, and in the limit of $T\ll q_{D}$, where
$q_{D}$ is the Debye wave number, it is given by \cite{Heiselberg:1993cr}\begin{equation}
\eta_{q}=\frac{1}{40\pi a}\left(\frac{2N_{q}}{\pi}\right)^{\frac{1}{3}}\alpha_{s}^{-\frac{5}{3}}\mu_{q}^{\frac{14}{3}}T^{-\frac{5}{3}}\label{eq:quark-shear}\end{equation}
where $\alpha_{s}=\frac{g^{2}}{4\pi}$ is the QCD coupling constant,
$a\simeq1.81$, $N_{q}=3$ and $\mu_{q}$ and $T$ are in units of
MeV. The temperature dependence arises again from a non-Fermi liquid
enhancement of the quark interaction. These expressions yield the
parameters in the parameterization of the shear viscosity eq. (\ref{eq:shear-parametrization})
as given on the right panel of table \ref{tab:weak-parameters}.

\section{R-mode time scales}

\subsection{General expressions}

The amplitude of the r-mode oscillations evolves with time dependence
$\mathrm{e}^{i\omega t-t/\tau}$, where $\omega$ is the real part
of the frequency of the r-mode and $1/\tau$ is the imaginary part
of the frequency. The latter describes both the exponential rise of
the r-mode driven by the Friedman-Schutz mechanism \cite{Friedman:1978hf}
and its decay due to viscous damping. We can decompose $1/\tau$ as
\[
\frac{1}{\tau\!\left(\Omega,T\right)}=\frac{1}{\tau_{G}\!\left(\Omega\right)}+\frac{1}{\tau_{B}\!\left(\Omega,T\right)}+\frac{1}{\tau_{S}\!\left(T\right)}\]
where $\tau_{G}$, $\tau_{B}$ and $\tau_{S}$ are gravitational radiation,
bulk viscosity and shear viscosity time scales, respectively. The
lowest of these individual time scales determines if the r-mode is
unstable or damped. The damping time for the individual mechanisms
is in general given by

\begin{equation}
\frac{1}{\tau_{i}}\equiv-\frac{1}{2E}\left(\frac{dE}{dt}\right)_{i}\end{equation}
and requires both the total energy of the r-mode 

\begin{equation}
E=\frac{1}{2}\alpha^{2}R^{4}\Omega^{2}\int_{0}^{R}\negmedspace\rho\!\left(r\right)\left(\frac{r}{R}\right)^{2m+2}dr\end{equation}
and the dissipated energy which is given by the corresponding part
of eq. (\ref{eq:dissipated-energy}) \cite{Lindblom:1999yk}. For
instance the dissipated energy eq. (\ref{eq:dissipated-energy}) due
to the bulk viscosity reads

\begin{align*}
\left(\frac{dE}{dt}\right)_{\zeta} & =-\kappa^{2}\Omega^{2}\int d^{3}x\left|\frac{\Delta\rho}{\rho}\right|^{2}\zeta\!\left(\left|\frac{\Delta\rho}{\rho}\right|^{2}\right)\end{align*}
where the dependence of the bulk viscosity on the conserved number
density fluctuation amplitude has been expressed in terms of the conserved
energy density. If the star consists of several shells consisting
of different forms of matter with different viscosity, as is the case
for hybrid stars, neutron stars with a high density core where direct
Urca reactions are allowed, etc., the integral consists of partial
integrals over the individual shells $s$ and the inverse viscous
damping times can be written as

\[
\frac{1}{\tau_{i}}=\sum_{s}\frac{1}{\tau_{i}^{\left(s\right)}}\]
Therefore, the contribution of the different shells enters additively.
We will in the following expressions suppress the explicit label $\left(s\right)$
but implicitly assume that the damping times consist of several contributions
when different layers are present. With these expressions the individual
r-mode time scales can be obtained. The time scale of the r-mode growth
due to gravitational wave emission is given by \cite{Lindblom:1998wf} 

\begin{equation}
\frac{1}{\tau_{G}}=-\frac{32\pi\left(m-1\right)^{2m}}{\left(\left(2m+1\right)!!\right)^{2}}\left(\frac{m+2}{m+1}\right)^{2m+2}\tilde{J_{m}}GMR^{2m}\Omega^{2m+2}\label{eq:gravitational-time-scale}\end{equation}
with the radial integral constant

\begin{equation}
\tilde{J_{m}}\equiv\frac{1}{MR^{2m}}\int_{0}^{R}\rho\!\left(r\right)r^{2m+2}dr\label{eq:J-tilde}\end{equation}

\subsubsection{Shear viscosity damping time}

The damping time of an r-mode with angular quantum number $m$ due
to shear viscosity is given by

\[
\frac{1}{\tau_{S}}=\frac{\left(m-1\right)\left(2m+1\right)}{\tilde{J}_{m}MR^{2m}}\int_{0}^{R}\eta\, r^{2m}dr\]
Using the parameterization eq. (\ref{eq:shear-parametrization}) this
can be written as

\begin{equation}
\frac{1}{\tau_{S}}=\frac{\left(m-1\right)\left(2m+1\right)\tilde{S}_{m}\Lambda_{QCD}^{3+\sigma}R}{\tilde{J}_{m}MT^{\sigma}}\label{eq:shear-viscosity-damping-time}\end{equation}
in terms of the dimensionless constant \begin{equation}
\tilde{S}_{m}\equiv\frac{1}{R^{2m+1}\Lambda_{QCD}^{3+\sigma}}\int_{R_{i}}^{R_{o}}\tilde{\eta}r^{2m}dr\label{eq:S-tilde}\end{equation}
where $R_{i}$ and $R_{o}$ are the inner and outer radius of the
corresponding shell, if there are several ones. This constant contains
the complete dependence on the particular microscopic processes. To
make this quantity dimensionless the generic scale $\Lambda_{QCD}$
has been introduced that is chosen as $\Lambda_{QCD}=1$ GeV for the
numeric values. For the fundamental $m=2$ r-mode the parameter $\tilde{S}$
is given for the different star models considered here in table \ref{tab:parameter-extrema-values}.

\subsubsection{Bulk viscosity damping time}

The bulk viscosity damping time is given by

\begin{align}
\frac{1}{\tau_{B}} & =\frac{\kappa^{2}}{\alpha^{2}\tilde{J}_{m}MR^{2}}\int d^{3}x\left|\frac{\Delta n}{\bar{n}}\right|^{2}\zeta\!\left(\left|\frac{\Delta n}{\bar{n}}\right|^{2}\right)\label{eq:bulk-damping-time}\end{align}
where the bulk viscosity is in general a function of the amplitude
\cite{Madsen:1992sx,Alford:2010gw}. As has been noted before the
strongly enhanced damping can provide a mechanism for the saturation
of the r-mode as will be discussed elsewhere. Here we restrict ourselves
to a study of the initial instability of small amplitude r-modes and
in the subthermal regime the viscosity eq. (\ref{eq:sub-viscosity})
is independent of the r-mode amplitude. In this case only the angular
integral over the density fluctuation enters and it is useful to define
the angular averaged form

\begin{align}
\delta\Sigma\!\left(r\right) & \equiv\frac{m+1}{2\alpha AR^{2}\Omega^{3}}\sqrt{\frac{\left(m+1\right)^{3}\left(2m+3\right)}{4m}}\left(\int d\Omega\left|\delta\sigma\right|^{2}\right)^{\frac{1}{2}}\nonumber \\
 & \xrightarrow[l.o.]{}\left(\frac{r}{R}\right)^{m+1}+\delta\Phi_{0}\end{align}
which reduces to the second line to leading order in the $\Omega$-expansion,
where the angular integral in eq. (\ref{eq:bulk-damping-time}) is
trivial due to the normalization of the spherical harmonics. The damping
time in the subthermal regime is then given by

\begin{equation}
\frac{1}{\tau_{B}^{<}}=\frac{16m}{\left(2m+3\right)\left(m+1\right)^{5}\kappa}\frac{R^{5}\Omega^{3}}{\tilde{J}_{m}M}\,{\cal T}_{m}^{<}\!\left(\frac{T^{\delta}}{\kappa\Omega}\right)\label{eq:subthermal-damping-time}\end{equation}
where the dependence on all local quantities, like the equation of
state, the weak rate, the density dependence of the particular star
model and its r-mode profile is contained in the function

\begin{align}
{\cal T}_{m}^{<}\!\left(b\right) & \equiv\frac{b}{R^{3}}\int_{R_{i}}^{R_{o}}\! dr\, r^{2}\frac{A^{2}C^{2}\tilde{\Gamma}}{1+\tilde{\Gamma}^{2}B^{2}b^{2}}\left(\delta\Sigma\!\left(r\right)\right)^{2}\label{eq:T-function}\end{align}
depending on a single external parameter and which has to be determined
numerically for a given star model. In the asymptotic limits the damping
time simplifies to

\begin{align}
\frac{1}{\tau_{B}^{<}} & \xrightarrow[f\ll1]{}\frac{16m}{\left(2m+3\right)\left(m+1\right)^{5}\kappa^{2}}\frac{\Lambda_{QCD}^{9-\delta}\tilde{V}_{m}R^{5}\Omega^{2}T^{\delta}}{\Lambda_{EW}^{4}\tilde{J}_{m}M}\label{eq:bulk-damping-time-low}\\
\frac{1}{\tau_{B}^{<}} & \xrightarrow[f\gg1]{}\frac{16m}{\left(2m+3\right)\left(m+1\right)^{5}}\frac{\Lambda_{EW}^{4}\Lambda_{QCD}^{\delta-1}\tilde{W}_{m}R^{5}\Omega^{4}}{\tilde{J}_{m}MT^{\delta}}\label{eq:bulk-damping-time-high}\end{align}
with the dimensionless constants

\begin{align}
\tilde{V}_{m} & \equiv\frac{\Lambda_{EW}^{4}}{R^{3}\Lambda_{QCD}^{9-\delta}}\int_{R_{i}}^{R_{o}}\! dr\, r^{2}A^{2}C^{2}\tilde{\Gamma}\,\left(\delta\Sigma\!\left(r\right)\right)^{2}\label{eq:V-tilde}\\
\tilde{W}_{m} & \equiv\frac{1}{R^{3}\Lambda_{EW}^{4}\Lambda_{QCD}^{\delta-1}}\int_{R_{i}}^{R_{o}}\! dr\, r^{2}\frac{A^{2}C^{2}}{\tilde{\Gamma}B^{2}}\left(\delta\Sigma\!\left(r\right)\right)^{2}\label{eq:W-tilde}\end{align}
Since the bulk viscosity originates from weak interactions here the
second normalization scale $\Lambda_{EW}$ is used with a generic
value $\Lambda_{EW}=100$ GeV. These normalization scales are only
introduced to obtain dimensionless constants of order one and drop
out of the final results for the damping times.

Finally, we want to stress that in these expressions for the r-mode
time scales the complete local dependence on the star profile, the
r-mode oscillation and the microscopic damping processes is contained
in the few constant $\tilde{J}$, $\tilde{S}$, $\tilde{V}$ and $\tilde{W}$,
but the dependence on the global parameters of the r-mode evolution
\cite{Owen:1998xg} $\Omega$ and $T$ is entirely explicit. These
constants include in particular also the complete dependence on the
non-trivial radial and angular dependence of the full next-to-leading
order expression for the r-mode\cite{Lindblom:1999yk}. Since to our
knowledge results for the bulk viscosity in the crust of neutron stars
are not yet available, we neglect the damping of the crust in our
numerical analysis. It has, however, been argued that the shear viscosity
of the crust could be crucial \cite{Bildsten:2000ApJ...529L..33B}
and a more detailed study of this issue is definitely desirable.

\begin{figure*}
\begin{minipage}[t]{0.5\textwidth}%
\includegraphics{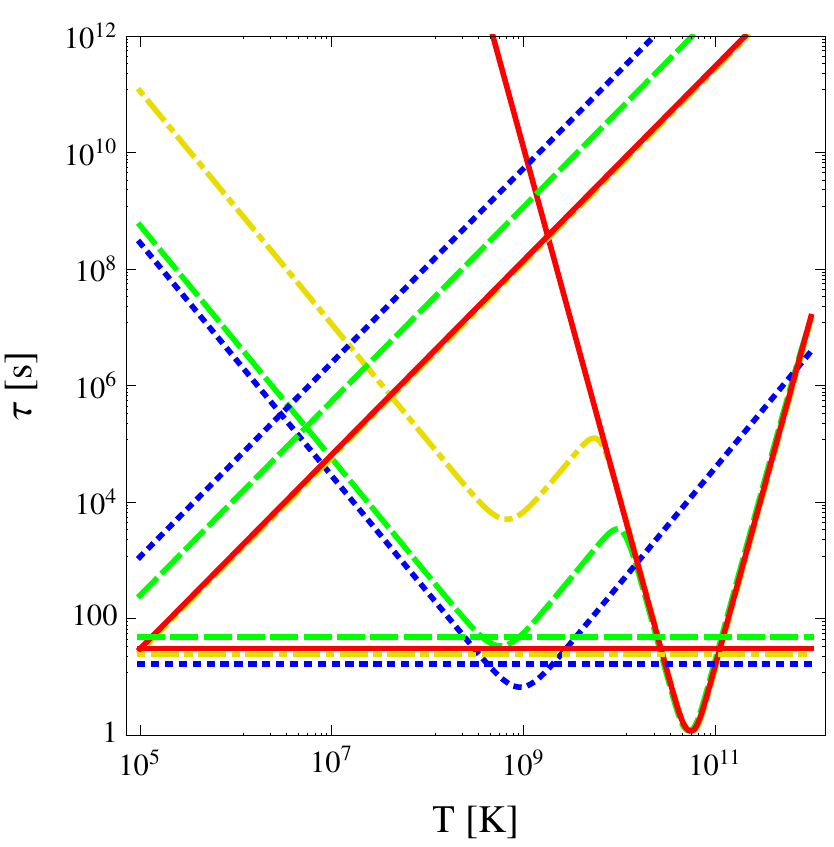}%
\end{minipage}%
\begin{minipage}[t]{0.5\textwidth}%
\includegraphics{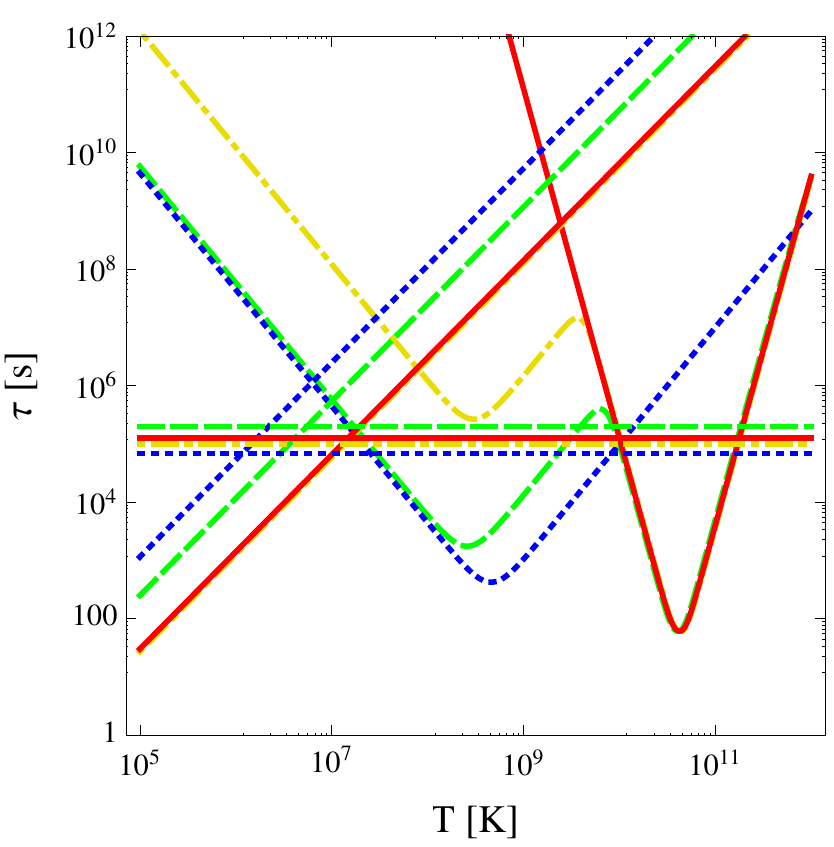}%
\end{minipage}

\caption{\label{fig:all-damping times}Damping times of the different $1.4\, M_{\odot}$
star models discussed in this work. Shown are a hadronic star with
modified Urca processes (solid), hybrid stars with a small (dot-dashed)
and large (dashed) quark core as well as a strange star (dotted).
The horizontal curves give the time scale $\tau_{G}$ associated to
the growth of the mode due to gravitational wave emission. The monotonically
increasing curves show the damping time $\tau_{S}$ due to shear viscosity
and the non-monotonic curves the damping time $\tau_{B}$ due to bulk
viscosity. \emph{Left panel:} Stars rotating at their Kepler frequency
$\Omega_{K}$. \emph{Right panel:} Same for stars rotating at $\Omega_{K}/4$.}

\end{figure*}

\subsection{Results for the considered star models}

Before we discuss the damping times for particular star models let
us point out a few generic properties of the damping times that follow
directly from the general expressions eqs. (\ref{eq:gravitational-time-scale}),
(\ref{eq:shear-viscosity-damping-time}) and (\ref{eq:subthermal-damping-time}).
The gravitational time scale is independent of temperature and decreases
strongly with frequency. The shear viscosity damping time increases
with temperature and is independent of frequency. Because of the resonant
behavior of the bulk viscosity, the corresponding damping time decreases
with temperature, has a minimum and increases again at large temperatures.
Furthermore, it also decreases with frequency, but slower than the
gravitational time scale.

A numeric solution requires the integration over the star profiles
to obtain the constants $\tilde{J}$, $\tilde{S}$ and the function
${\cal T}^{<}$, respectively the constants $\tilde{V}$ and $\tilde{W}$
describing its asymptotic behavior. The constants are given for the
fundamental $m=2$ r-mode of the various star models discussed in
this work, and where applicable also for their different shells, in
table \ref{tab:parameter-extrema-values}. For the bulk viscosity
of a $1.4\, M_{\odot}$ neutron star we compute the APR result employing
the proper susceptibilities for interacting matter, as well the result
when the susceptibilities are evaluated in the idealized case of a
hadronic gas, as has been done previously \cite{Cutler:1990,Sawyer:1989dp}.
As can be seen the parameters in the interacting case are more than
an order of magnitude larger, owing to the larger viscosity \cite{Alford:2010gw},
which leads to a correspondingly smaller damping time. In the following
we will consider only the proper interacting form unless otherwise
noted.

\begin{table*}
\begin{tabular}{|c|c|c|c|c|c|c|c|c|c|}
\hline 
star model & shell & $\tilde{J}$ & $\tilde{S}$ & $\tilde{V}$ & $\tilde{W}$ & $T_{min}$$\left[K\right]$ & $\Omega_{min}$$\left[Hz\right]$ & $T_{max}$$\left[K\right]$ & $\Omega_{max}$$\left[Hz\right]$\tabularnewline
\hline 
NS $1.4\, M_{\odot}$ & core & $1.81\times10^{-2}$ & $7.68\times10^{-5}$ & $1.31\times10^{-3}$ & $1.61\times10^{-6}$ & $3.49\times10^{9}$ & $371$ & $-$ & $-$\tabularnewline
\cline{1-1} \cline{4-10} 
NS $1.4\, M_{\odot}$ gas &  &  & $4.32\times10^{-6}$ & $1.28\times10^{-4}$ & $1.76\times10^{-7}$ & $3.70\times10^{9}$ & $226$ & $-$ & $-$\tabularnewline
\cline{1-1} \cline{3-10} 
NS $2.0\, M_{\odot}$ &  & $2.05\times10^{-2}$ & $2.25\times10^{-4}$ & $1.16\times10^{-3}$ & $1.72\times10^{-6}$ & $4.19\times10^{9}$ & $368$ & $-$ & $-$\tabularnewline
\hline
NS $2.21\, M_{\odot}$ & d.U. core & $2.02\times10^{-2}$ & $5.05\times10^{-4}$ & $1.16\times10^{-8}$ & $7.11\times10^{-7}$ & $2.03\times10^{9}$ & $493$ & $-$ & $-$\tabularnewline
\cline{2-2} \cline{5-6} 
 & m.U. core &  &  & $9.34\times10^{-4}$ & $1.55\times10^{-6}$ &  &  &  & \tabularnewline
\hline
SS eq. (\ref{eq:quark-eos-model}) & all & $\frac{3}{28\pi}$ & $\frac{\hat{\eta}\mu_{q}^{14/3}}{5\Lambda_{QCD}^{14/3}\alpha_{s}^{5/3}}$ & $\frac{\Lambda_{EW}^{4}\hat{\Gamma}m_{s}^{4}\mu_{q}^{3}}{9\Lambda_{QCD}^{7}(1-c)^{2}}$ & $\frac{m_{s}^{4}}{4\pi^{4}\Lambda_{EW}^{4}\Lambda_{QCD}\hat{\Gamma}\mu_{q}^{3}}$ & eq. (\ref{eq:SS-minimum-temperature}) & eq. (\ref{eq:SS-minimum-frequency}) & eq. (\ref{eq:SS-maximum-temperature}) & eq. (\ref{eq:SS-maximum-frequency})\tabularnewline
\cline{1-1} \cline{3-10} 
SS $1.4\, M_{\odot}$ &  & $3.08\times10^{-2}$ & $3.49\times10^{-6}$ & $3.53\times10^{-10}$ & $0.191$ & $7.86\times10^{6}$ & $1020$ & $1.01\times10^{9}$ & $8340$\tabularnewline
\cline{1-1} \cline{3-10} 
SS $2.0\, M_{\odot}$ &  & $2.65\times10^{-2}$ & $4.45\times10^{-6}$ & $3.38\times10^{-10}$ & $0.157$ & $8.58\times10^{6}$ & $955$ & $8.07\times10^{8}$ & $6600$\tabularnewline
\hline
HS $1.4\, M_{\odot}$ L & quark core & $1.93\times10^{-2}$ & $2.19\times10^{-6}$ & $1.38\times10^{-10}$ & $1.03\times10^{-2}$ & $9.87\times10^{6}$ & $1170$ & $5.83\times10^{8}$ & $7270^{*}$\tabularnewline
\cline{2-2} \cline{4-10} 
 & hadr. core &  & $6.72\times10^{-6}$ & $1.40\times10^{-3}$ & $1.51\times10^{-6}$ & $6.88\times10^{9}$ & $1040^{\dagger}$ & $-$ & $-$\tabularnewline
\hline
HS $1.4\, M_{\odot}$ S & quark core & $1.68\times10^{-2}$ & $2.88\times10^{-7}$ & $7.96\times10^{-13}$ & $6.85\times10^{-5}$ & $1.01\times10^{8}$ & $1070$ & $4.35\times10^{8}$ & $3690^{*}$\tabularnewline
\cline{2-2} \cline{4-10} 
 & hadr. core &  & $3.29\times10^{-6}$ & $1.25\times10^{-3}$ & $1.56\times10^{-6}$ & $3.61\times10^{9}$ & $394^{\diamond}$ & $-$ & $-$\tabularnewline
\hline
HS $2.0\, M_{\odot}$ & quark core & $2.00\times10^{-2}$ & $5.25\times10^{-6}$ & $3.76\times10^{-10}$ & $2.34\times10^{-2}$ & $7.73\times10^{6}$ & $1080$ & $4.77\times10^{8}$ & $6310^{*}$\tabularnewline
\cline{2-2} \cline{4-10} 
 & hadr. core &  & $5.24\times10^{-6}$ & $1.07\times10^{-3}$ & $1.32\times10^{-6}$ & $7.65\times10^{9}$ & $925^{\dagger}$ & $-$ & $-$\tabularnewline
\hline
\end{tabular}

\caption{\label{tab:parameter-extrema-values}Radial integral parameters and
characteristic points of the instability region of a $m\!=\!2$ r-mode
for the star models considered in this work. The constant $\tilde{J}$,
$\tilde{S}$, $\tilde{V}$ and $\tilde{W}$ are given by eqs. (\ref{eq:J-tilde}),
(\ref{eq:S-tilde}), (\ref{eq:V-tilde}) and (\ref{eq:W-tilde}) using
the generic normalization scales $\Lambda_{QCD}=1$ GeV and $\Lambda_{EW}=100$
GeV. The temperatures and frequencies are obtained with the analytic
expressions for the minima eqs. (\ref{eq:minimum-frequency}), (\ref{eq:minimum-temperature}),
(\ref{eq:second-minimum-frequency}) and (\ref{eq:second-minimum-temperature})
and for the maxima eqs. (\ref{eq:maximum-frequency}) and (\ref{eq:maximum-temperature}).
The expressions for a generic strange star (or quark core) in terms
of the parameters of the quark model equation of state eq. (\ref{eq:quark-eos-model}),
using the constants $\hat{\Gamma}$ and $\hat{\eta}$ defined in eq.
(\ref{eq:quark-viscosity-prefactors}) are given as well. $\left(*\right)$
These values deviate significantly from the actual results due to
the inappropriate approximation of constant radial profiles, whereas
this idealization entirely fails for hadronic parts. $\left(\dagger\right)$
The second minimum arises from the competition of the bulk viscosity
damping in the quark and the hadronic shell. $\left(\diamond\right)$
The second minimum arises from the competition of bulk and shear viscosity
damping in the hadronic shell.}

\end{table*}

In the left panel of fig.~\ref{fig:all-damping times} the numeric
solution for the different time scales is shown as a function of temperature
for $1.4\, M_{\odot}$ stars of the different classes considered here
rotating at their Kepler frequency $\Omega_{K}$. Shown are gravitational
time scales (horizontal lines), the shear viscosity damping times
(monotonically increasing curves) and bulk viscosity damping times
(non-monotonic curves). The solid lines denote the neutron star model,
the dotted lines a strange star and the dashed and dot-dashed lines
show hybrid star models with a small and large quark core, respectively.
Whereas damping due to shear viscosity dominates for strange and hybrid
stars for $T\lesssim10^{7}$ K and for neutron stars even for $T\lesssim10^{9}$
K, the bulk viscosity damping time of the neutron and the strange
star feature the generic resonant form, where the minima are around
$10^{9}$ K and $10^{11}$ K, respectively, and the higher gradients
in case of the neutron star arise from the higher power of $\delta$
(tab. \ref{tab:weak-parameters}). It is clear that as a function
of the core size the family of hybrid curves interpolates continuously
between the two uniform star models. Correspondingly, for hybrid stars
the contribution of the quark core dominates at low temperature, whereas
the hadronic shell dominates at higher temperatures, which leads to
a curve with two minima. For all star models at $\Omega_{K}$the gravitational
time scale is the lowest over a range of temperatures.

The right panel of fig.~\ref{fig:all-damping times} shows the same
plot at the lower frequency $\Omega_{K}/4$. As is clear from the
general discussion the shear curves are unchanged whereas the gravitational
curves and the bulk curves are shifted upwards compared to the left
panel of fig.~\ref{fig:all-damping times}. Since the gravitational
time scale moves upwards faster than the bulk viscosity curve, the
instability region shrinks from both sides as the frequency is lowered.
For hybrid stars with a large quark core the gravitational time scale
can move above the minimum of the bulk viscosity arising from the
quark core as the frequency is lowered and the instability region
is divided by a stability window. Yet, for a hybrid star with a sufficiently
small quark core the minimum can move above the shear curve before
the gravitational time scale can overtake it. In this case there would
be no signature of the quark core and the instability regions of such
a hybrid star and of the corresponding neutron star would look basically
indistinguishable%
\footnote{The fact that the volume of the neutron shell is slightly smaller
is not significant for the damping times that vary over many orders
of magnitude, so that the dashed curves in fig.~\ref{fig:all-damping times}
are invisible underneath the solid neutron star curves.%
}. This is nearly the case for the small core hybrid star model denoted
by the dot-dashed curve as will be discussed below.

\section{R-mode instability regions}

The boundary of the instability region is given by the condition $\left|\tau_{G}\right|=\left|\tau_{V}\right|$,
where $\tau_{V}$ is the viscous damping time. This reads in terms
of the individual contributions

\begin{equation}
\frac{1}{\tau_{G}}+\sum_{s}\left(\frac{1}{\tau_{S}^{\left(s\right)}}+\frac{1}{\tau_{B}^{\left(s\right)}}\right)=0\label{eq:boundary-condition}\end{equation}
Since at the boundary the time scale of the gravitational instability
is identical to the viscous time scale, the r-mode amplitude neither
increases nor decreases. If the viscous damping time is shorter modes
induced by external perturbations will quickly be damped away, whereas
in the opposite case they are unstable and will initially grow. As
will be shown in a second article the increase in the bulk viscosity
at large amplitude can eventually saturate the r-mode, but here we
will limit ourselves to the small amplitude regime and analyze the
regions where small amplitude modes are unstable. In general eq. (\ref{eq:boundary-condition})
has no analytic solution and has to be solved numerically. However
there are various limiting cases for which analytic solutions exist
and we will study them below.

\subsection{Analytic expressions}

The boundary of the instability region eq. (\ref{eq:boundary-condition})
can generally not be found analytically due to the occurrence of several
terms with non-trivial temperature and frequency dependencies. Since
the viscous damping times vary extremely strongly with temperature
and frequency (fig.~\ref{fig:all-damping times}) in general there
is over nearly the whole parameter space one component that clearly
dominates the others in the sum. In such a case the equation can easily
be solved analytically and yields analytic expressions for the different
segments of the instability region. Although the numeric solution
of eq. (\ref{eq:boundary-condition}) is straightforward, these analytic
expressions reveal the general dependence on the various unknown model
parameters entering the r-mode analysis and therefore provide a measure
for the uncertainty of these results in a situation where most properties
of dense strongly interacting are basically unknown. Here we give
expressions for the fundamental $m\!=\!2$ r-mode and the classes
of stars discussed in this work and defer the general results to appendix
\ref{sec:Semi-analytic-boundary}.

\subsubsection{Low temperature boundary}

As is clear from the analytic expressions for the damping times and
confirmed by fig.~\ref{fig:all-damping times} at low temperatures
shear viscosity damping dominates. Furthermore, in the case of hybrid
stars the damping due to hadronic shear viscosity dominates over the
quark core. Comparing it with the gravitational time scale yields

\begin{align*}
\Omega\left(T\right) & \xrightarrow[T\ll T_{min}]{}1.12\frac{\tilde{S}^{\frac{1}{6}}\Lambda_{QCD}^{\frac{7}{9}}}{\tilde{J}^{\frac{1}{3}}G^{\frac{1}{6}}M^{\frac{1}{3}}R^{\frac{1}{2}}}T^{-\frac{5}{18}}\quad\left(\mathrm{NS}\,\mathrm{and}\,\mathrm{SS}\right)\end{align*}
where the irrational power arises from the Landau damping of the corresponding
interactions that induce shear viscosity. Interestingly this expression
depends only mildly on the constant $\tilde{S}$ that encodes the
microscopic interaction. This segment of the boundary is important
for old stars in binary systems that are spun up by accretion from
a companion star and enter the instability region from below

\subsubsection{Minimum of the instability region}

 Since the shear viscosity damping time monotonically increases with
temperature whereas the bulk viscosity damping time monotonically
decreases at low temperature there is a minimum of the instability
boundary given by $\tau_{B}=\tau_{S}$. Due to the non-linear frequency
dependence of the second order damping times an exact analytic solution
is not possible. However, since these minima are located at frequencies
far below the Kepler frequency it is according to fig. \ref{fig:oscillation frequency}
a very good approximation to use the leading order frequency connection
$\kappa=\kappa_{0}$ in which case an analytic solution is possible.
For neutron stars it reads

\begin{align}
\Omega_{min}^{\left(NS\right)} & \approx1.06\frac{\Lambda_{QCD}^{\frac{99}{128}}\tilde{S}^{\frac{9}{64}}\tilde{V}^{\frac{5}{128}}}{R^{\frac{49}{128}}\tilde{J}^{\frac{23}{64}}G^{\frac{23}{128}}M^{\frac{23}{64}}\Lambda_{EW}^{\frac{5}{32}}}\label{eq:NS-minimum-frequency}\\
T_{min}^{\left(NS\right)} & \approx1.89\frac{\tilde{S}^{\frac{3}{32}}\Lambda_{QCD}^{\frac{1}{64}}\Lambda_{EW}^{\frac{9}{16}}G^{\frac{3}{64}}\tilde{J}^{\frac{3}{32}}M^{\frac{3}{32}}}{\tilde{V}^{\frac{9}{64}}R^{\frac{27}{64}}}\label{eq:NS-minimum-temperature}\end{align}
Note, the appearance of surprisingly low powers in these expressions.
In particular, due to the arising $5/128$ power a change of $\tilde{V}$
by an order of magnitude results only in a mild deviation of the minimum
frequency of less than $10$\%, and even a very drastic change by
three orders of magnitude does not change the result by more than
$30$\%. Whereas the viscosity constants eqs. (\ref{eq:V-tilde})
and (\ref{eq:W-tilde}) can vary by many orders of magnitude for different
classes of stars with different transport processes as can be seen
from tab. \ref{tab:parameter-extrema-values}, they are generically
of similar order of magnitude within a given class due to the identical
parametric dependence on the microscopic physics. Recall, that the
minimum of the instability region is of particular importance for
the r-mode analysis since it determines to what frequency r-modes
can spin down a star. It is needless to say that such an insensitivity
of the minimum frequency on the microscopic physics is more than welcome
in the present situation where there are still huge uncertainties
on the underlying equation of state and the transport coefficients
of dense matter. This presents one of the main results of this article. 

In the general case studied in appendix \ref{sec:Semi-analytic-boundary}
the $\left(2m\left(\delta+\sigma\right)+2\delta\right)/\sigma$-th
root of the bulk viscosity constants $\tilde{V}_{m}$ arises in the
expression for the minimum frequency eq. (\ref{eq:minimum-frequency}).
For short-range (Fermi liquid) interactions the shear viscosity exponent
is generically $\sigma=2$ whereas it reduces to $\sigma=5/3$ when
long-ranged, only Landau-damped (non-Fermi liquid) gauge interactions
are present. Since the rate of the weak interactions vanishes in equilibrium
and requires phase space both for initial and final state particles
in general $\delta\geq2$. A similar argument should hold for processes
mediated by Goldstone bosons in color superconducting phases. Therefore
the above root is for all multipoles and corresponding processes
higher than $11$ so that change of the viscosity constants $\tilde{V}_{m}$
by an order of magnitude still changes the minimum only by at most
$\sim20$\%, making it very insensitive to them. This extends the
finding obtained in an explicit comparison of various different neutron
star models in \cite{Lindblom:1998wf} to general forms of matter
and damping processes. This insensitivity is particularly interesting
since these constants contain the complete dependence on the detailed
second order r-mode profiles, so that the minimum frequency is hardly
affected by the second order effects. This observation had already
been made for the particular neutron star model studied in \cite{Lindblom:1999yk}.
A similar statement holds for the dependence on the shear viscosity
constants $\tilde{S}_{m}$, where the corresponding root is at least
of $6$th order reached in the limit $\delta\gg\sigma$. The temperature
of the minimum is more sensitive to these constants, as likewise observed
in \cite{Lindblom:1998wf,Lindblom:1999yk}, since it only involves
the $\left(\sigma+\delta\right)$-th root.

Finally we give the explicit expression for the minimum frequency
of strange stars with the general quark model equation of state eq.
(\ref{eq:quark-eos-model}). Here the parameter dependence of the
quark matter viscosity coefficients given in table \ref{tab:weak-parameters}
can be factored out according to

\begin{equation}
\tilde{\Gamma}^{\left(q\right)}\equiv\hat{\Gamma}\mu_{q}^{5}\quad,\tilde{\eta}^{\left(q\right)}\equiv\hat{\eta}\alpha_{s}^{-5/3}\mu_{q}^{14/3}\label{eq:quark-viscosity-prefactors}\end{equation}
where $\hat{\Gamma}$ and $\hat{\eta}$ are pure constants, so that
the minimum is located at

\begin{align}
\Omega_{min}^{\left(SS\right)} & \approx2.23\frac{\hat{\Gamma}^{\frac{5}{56}}\hat{\eta}^{\frac{3}{28}}m_{s}^{\frac{5}{14}}\mu_{q}^{\frac{43}{56}}}{G^{\frac{11}{56}}(1-c)^{\frac{5}{28}}\alpha_{s}^{\frac{5}{28}}R^{\frac{13}{56}}M^{\frac{11}{28}}}\label{eq:SS-minimum-frequency}\\
T_{min}^{\left(SS\right)} & \approx2.78\frac{\hat{\eta}^{\frac{3}{14}}G^{\frac{3}{28}}\mu_{q}^{\frac{1}{28}}\left(1-c\right)^{\frac{9}{14}}M^{\frac{3}{14}}}{\hat{\Gamma}^{\frac{9}{28}}\alpha_{s}^{\frac{5}{14}}m_{s}^{\frac{9}{7}}R^{\frac{27}{28}}}\label{eq:SS-minimum-temperature}\end{align}
Assuming that eq. (\ref{eq:quark-eos-model}) gives an estimate for
the uncertainty in the unknown quark matter equation of state and
estimating the uncertainty in the quadratic and quartic parameters
$m_{s}$ and $1-c$ each by a factor of two and that in the strong
coupling $\alpha_{s}$ generously by an order of magnitude, the minimum
could vary here by roughly a factor of two owing to the uncertainty
in the unknown microscopic dynamics. As noted before the quark model
equation of state is also valid for color superconducting phases \cite{Alford:2004pf}.
In this case the gap reduces the parameter $m_{s}$ which would lower
the minimum frequency and enlarge the instability region.

\subsubsection{Intermediate boundary}

Above the minimum the damping is dominated by the bulk viscosity and
below its maximum it can be approximated by the asymptotic low temperature
form eq. (\ref{eq:bulk-damping-time-low}). The semi-analytic result
to next-to leading order in $\Omega$ is given in the appendix, but
because of the small effect of the next-to-leading order corrections
of the oscillation frequency on the instability regions, observed
above, we can simply neglect these and obtain \begin{align*}
 & \Omega\left(T\right)\xrightarrow[T\ll T_{max}]{T\gg T_{min}}0.360\frac{\tilde{V}^{\frac{1}{4}}\Lambda_{QCD}^{\frac{9}{4}}R^{\frac{1}{4}}}{\tilde{J}^{\frac{1}{2}}\Lambda_{EW}G^{\frac{1}{4}}M^{\frac{1}{2}}}\cdot\left\{ \begin{array}{cc}
\!\!\left(\frac{T}{\Lambda_{QCD}}\right)^{\frac{3}{2}} & \left(\mathrm{NS}\right)\\
\!\!\left(\frac{T}{\Lambda_{QCD}}\right)^{\frac{1}{2}} & \left(\mathrm{SS}\right)\end{array}\right.\end{align*}
This part of the instability boundary is important for the spin down
of young compact stars since it determines where they hit the instability
region during their initial fast cooling phase.

\subsubsection{Maximum of the instability region for strange stars}

Due to the resonant form of the bulk viscosity eq. (\ref{eq:sub-viscosity})
whose maximum eq. (\ref{eq:max-viscosity}) translates into a minimum
of the corresponding damping time, the instability region features
a maximum if the corresponding frequency is below the Kepler frequency
or otherwise splits into two parts. Since there is no analytic expression
for the damping time in the vicinity of the maximum there is no general
expression for the maximum. Yet, for hadronic matter the corresponding
maximum frequency is usually above the Kepler frequency anyway so
it is not relevant for physical applications. In contrast, for strange
stars, where the maximum is generally below the Kepler frequency and
important for the r-mode analysis, the density profile varies only
mildly and can be approximated by a uniform density. In this case
a general analytic expression for the maximum of the instability region
is given in eqs. (\ref{eq:maximum-frequency}, \ref{eq:maximum-temperature})
in appendix \ref{sec:Semi-analytic-boundary} and reduces using the
susceptibilities in tab. \ref{tab:strong-parameters} to 

\begin{align}
\Omega_{max}^{\left(SS\right)} & \approx0.434\frac{m_{s}^{\frac{4}{3}}R^{\frac{1}{3}}}{\left(1-c\right)^{\frac{1}{3}}G^{\frac{1}{3}}M^{\frac{2}{3}}}\label{eq:SS-maximum-frequency}\\
T_{max}^{\left(SS\right)} & \approx0.210\frac{\left(1-c\right)^{\frac{1}{3}}m_{s}^{\frac{2}{3}}R^{\frac{1}{6}}}{\hat{\Gamma}^{\frac{1}{2}}G^{\frac{1}{6}}\mu_{q}^{\frac{3}{2}}M^{\frac{1}{3}}}\label{eq:SS-maximum-temperature}\end{align}
where the tiny second order corrections to the oscillation frequency,
cf. fig.~\ref{fig:oscillation frequency}, were neglected in this
case. Note that analogous to the expression for the maximum of the
viscosity of quark matter given in \cite{Alford:2010gw} the chemical
potential drops out and therefore there is no ambiguity where to evaluate
the susceptibilities in case the density distribution of the considered
star model is not entirely constant. This gives for the dimensionless
ratio

\begin{equation}
\frac{\Omega_{max}^{\left(SS\right)}}{\Omega_{K}}\approx1.03\frac{m_{150}^{\frac{4}{3}}R_{10}^{\frac{11}{6}}}{\left(1-c\right)^{\frac{1}{3}}M_{1.4}^{\frac{7}{6}}}\label{eq:maximum-ratio}\end{equation}
where $m_{150}$, $M_{1.4}$ and $R_{10}$ are the effective strange
quark mass in units of 150 MeV, the stars mass in units of $1.4\, M_{\odot}$
and the radius in units of $10$ km, respectively. 

Whereas the dependence on the parameter $c$, which in the perturbative
regime is positive and thereby increases the maximum frequency, is
comparatively mild, the dependence on the effective strange quark
mass is significant and can within the probable uncertainty region
$100\,\mathrm{MeV}\lesssim m_{s}\lesssim200\,\mathrm{MeV}$ strongly
change the position of the maximum frequency from values considerably
below to values above the Kepler frequency. For superconducting matter
the gap can reduce the effective parameter $m_{s}$. Whereas the maximum
frequency of massive stars can be significantly lower, the radii of
sufficiently massive strange stars do not vary much with mass along
a mass-radius curve so that the dependence on the radius is mild despite
the arising high power. To judge the implicit dependence of the masses
and radii on the equation of state it is useful to recall that the
solution of the TOV equations for a uniform density star exhibits
scaling with the bag constant $R/R_{0},M/M_{0}\sim\sqrt{B_{0}/B}$
\cite{Glendenning}. Despite the uniform density approximation, the
analytic result eq. (\ref{eq:maximum-ratio}) for the maximum, given
in tab. \ref{tab:parameter-extrema-values}, agrees nicely with the
numeric results for the considered strange star models. The above
expression can also be applied to the corresponding analysis of Jaikumar,
et. al. \cite{Jaikumar:2008kh}. Using their lower strange quark mass
$m_{s}=100$ MeV as well as their slightly smaller radius and taking
into account that these authors erroneously used the angular velocity
in the inertial frame to evaluate the bulk viscosity yields $\Omega_{max}^{\left(SS\right)}/\Omega_{K}\approx0.47$
in very good agreement with the plot of the exclusion region shown
in their fig.~4. When employing the proper angular velocity in the
rotating frame, eq. (\ref{eq:maximum-ratio}), gives instead the corrected
result $\Omega_{max}^{\left(SS\right)}/\Omega_{K}\approx0.59$.

The analytic results in appendix \ref{sec:Semi-analytic-boundary}
to some extent also apply to the maxima of the instability regions
of hybrid stars but in this case the agreement is less precise since
the density profile of their quark core features a more pronounced
radial dependence.

\subsubsection{High temperature boundary}

For completeness we also give the high temperature part of the boundary. 

\[
\Omega\left(T\right)\xrightarrow[T\gg T_{max}]{}8.66\times10^{-2}\frac{\tilde{W}^{\frac{1}{2}}\Lambda_{EW}^{2}R^{\frac{1}{2}}}{\tilde{J}\Lambda_{QCD}^{\frac{1}{2}}G^{\frac{1}{2}}M}\cdot\left\{ \begin{array}{cc}
\!\!\left(\frac{\Lambda_{QCD}}{T}\right)^{3} & \left(\mathrm{NS}\right)\\
\!\!\frac{\Lambda_{QCD}}{T} & \left(\mathrm{SS}\right)\end{array}\right.\]
Due to the resonant behavior of the bulk viscosity which decreases
again at large temperatures, the r-mode is unstable for all forms
of matter at large temperatures above the corresponding boundary of
the shell with the largest resonance temperature eq. (\ref{eq:max-viscosity}).
This part could be less relevant for the r-mode evolution since stars
cool very fast initially and could leave this region before the r-mode
can develop.

\subsubsection{General properties of the analytic analysis}

Finally let us note a few generic properties of these analytic results.
Since the mass appears in the denominator in all these expressions,
an increase of the mass of the star increases the instability region
and moves it uniformly downward and slightly to the right in a $T-\Omega$-plot.
The dependence on the radius is less uniform, but in general these
expressions are less sensitive to the radius than to the mass of the
star. Furthermore for a given equation of state the radius of a star
varies very little for masses of $1$ to $2\, M_{\odot}$. As already
noted, the dependence of the minima of the instability region on the
viscosity parameters is extremely mild. Due to the arising small powers
$O\left(1/10\right)$ even a large variation of the viscosity parameters
by three orders of magnitude within the set of possible equations
of state for a given class of stars, like e.g. pure neutron stars
would not change the minimum by more than a factor of two. In contrast
the dependence of the maxima on the viscosity parameters is far more
pronounced and as discussed above changes by more than a factor of
two are here easily possible.

\subsection{Numeric results}

\begin{figure*}
\begin{minipage}[t]{0.5\textwidth}%
\includegraphics{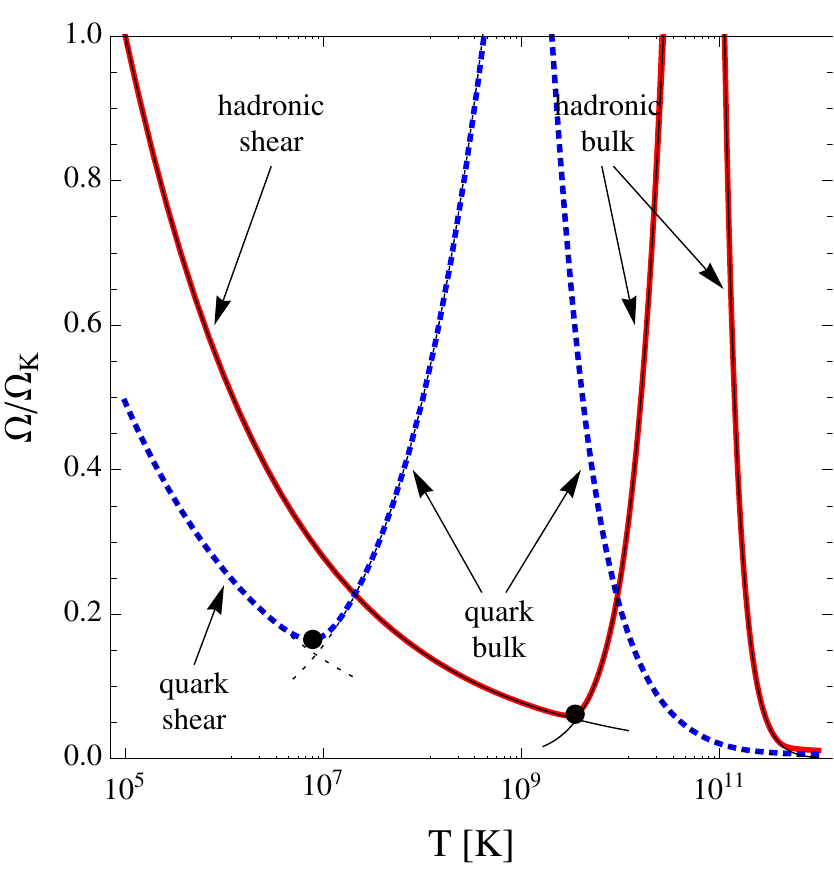}%
\end{minipage}%
\begin{minipage}[t]{0.5\textwidth}%
\includegraphics{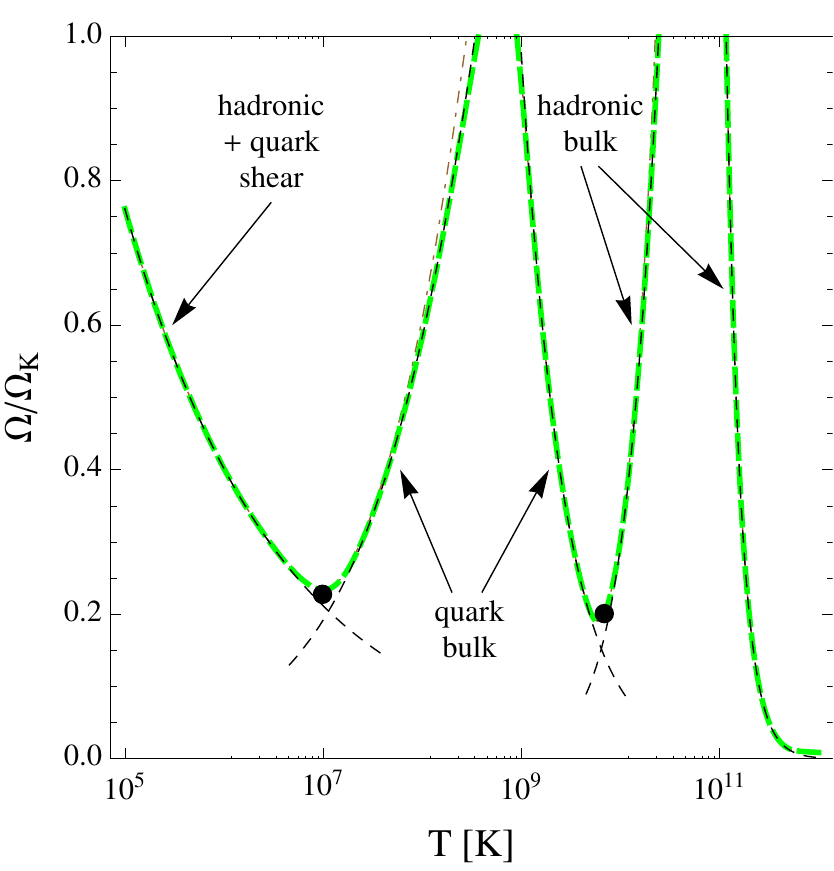}%
\end{minipage}

\caption{\label{fig:instability-regions-analytic}Comparison of the numeric
results for the instability region (thick lines) with the various
approximate semi-analytic expressions (thin lines and blobs) presented
in the text for the different classes of compacts stars. The dots
present the results for the extrema and the thin lines which are valid
away from the extrema represent the corresponding analytic results
taking into account only the contribution from the dominant process
and shell in the respective region. \emph{Left panel:} Instability
regions for the $1.4\, M_{\odot}$ neutron star (solid) and strange
star model (dotted). \emph{Right panel:} Same for the $1.4\, M_{\odot}$
hybrid star model. The thin dot-dashed curve on the right panel which
deviates slightly close to the maximum is the leading order result
without the frequency corrections in fig.~\ref{fig:oscillation frequency},
whereas the corresponding curves are indistinguishable on the left
panel.  Note that here in the following where the ratio $\Omega/\Omega_{K}$
is plotted this ratio is taken with the respective Kepler angular
frequency for each star model, see table \ref{tab:star-models}, so
that the same value of $\Omega/\Omega_{K}$ corresponds to a different
value of $\Omega$ for different curves. }

\end{figure*}

Let us now discuss the numeric results for the exclusion regions and
compare them to the semi-analytic expressions obtained in the last
section in order to assess the quality of the latter. Fig.~\ref{fig:instability-regions-analytic}
compares the expressions for particular segments of the instability
region and for the extrema to the numeric solution. The left panel
shows the results for a standard $1.4\, M_{\odot}$ neutron star model
(solid) and a corresponding strange star model (dotted). The thick
curves show the numeric results and the thin curves the analytic approximations
and the dots denote the minima. Since we neglect the contribution
from the neutron star crust there is only a single shell in each case.
As can be seen, except for the regions around the extrema, the analytic
expressions for the different segments approximate the boundary of
the instability region extremely well and are mostly hidden underneath
the numeric curves. Yet, the extrema are in turn well described by
the corresponding analytic expressions. As noted in \cite{Madsen:1999ci,Andersson:2000mf,Jaikumar:2008kh},
due to the resonant form of the bulk viscosity there is a stability
window around $10^{9}$ K where strange stars are not unstable against
r-modes up to large frequencies. For the considered strange star model
the maximum of the instability region is above the Kepler frequency
so that two separate regions appear, but as discussed before the position
of the maximum depends on the particular microscopic parameters. Due
to the same qualitative resonant structure of the bulk viscosity there
is also a second instability region at high temperatures $\gtrsim10^{11}$
K for neutron stars that has previously been neglected due to the
employed low temperature approximation to the hadronic bulk viscosity
\cite{Cutler:1990,Sawyer:1989dp}. Although the r-mode is initially
unstable the naive expectation is that this should be irrelevant for
the spin-down evolution since the star cools extremely fast at such
high temperatures and might leave this region before r-modes can develop.
Yet, since the interior of the star is initially opaque to neutrinos
and the cooling is delayed this is not entirely clear and requires
further study.

On the right panel of fig.~\ref{fig:instability-regions-analytic}
the comparison for the $1.4\, M_{\odot}$ hybrid star model is shown.
Here there are contributions from the different layers, but strikingly
the analytic approximation works even in this more complicated case.
This plot shows nicely the generic structure of the boundary of the
instability region, discussed in the previous subsection, when several
shells are present that feature qualitatively different damping mechanisms.
Around its resonant temperature the bulk viscosity damping mechanism
of a given shell in general clearly dominates over those of the other
shells. If the resonant temperatures of the different bulk viscosities
are sufficiently separated there are several maxima, that define corresponding
stability windows. Below each of these maxima there is a minimum where
a dominant mechanism is replaced by the next. However, in case two
resonant temperatures are too close or a shell is too small to have
a sizable impact, individual stability windows can be fully or partly
washed out. We discuss such cases below. 

\begin{figure}
\includegraphics{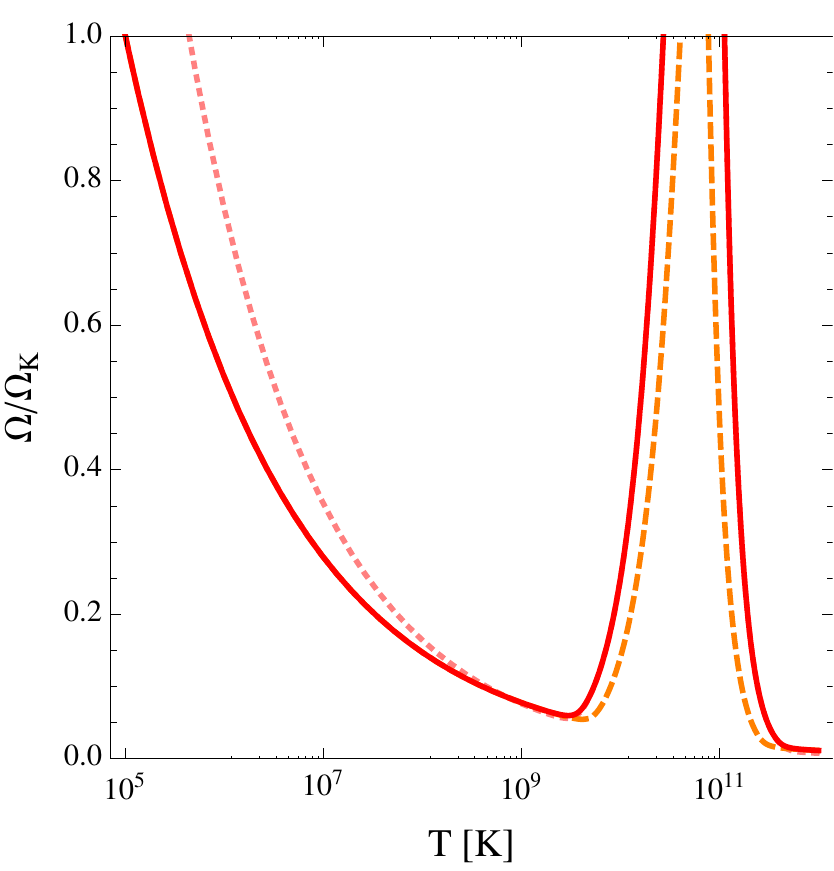}

\caption{\label{fig:instability-regions-neutron-modification}Modification
of the instability region of a $1.4\, M_{\odot}$ neutron star due
to improved approximations for the microscopic transport properties.
The solid curve shows the standard neutron star model with shear viscosity
due to the dominant Landau-damped lepton scattering and bulk viscosity
based on the proper susceptibilities for interacting matter (as all
other neutron star results in this work). The dashed curve shows the
result when using the fit given in \cite{Cutler:1990} to the density
data $\lesssim n_{0}$ for the shear viscosity from hadron-hadron
scattering obtained in \cite{Flowers:1976ApJ...206..218F}. The dotted
line shows the result when employing the previously used expression
neglecting interactions to the susceptibilities contributing to the
bulk viscosity \cite{Cutler:1990,Sawyer:1989dp}.}

\end{figure}

Fig.~\ref{fig:instability-regions-neutron-modification} shows the
exclusion region for the standard $1.4\, M_{\odot}$ neutron star
model compared to approximations used previously in the literature.
The solid curve shows our new result with shear viscosity due to the
dominant Landau-damped lepton scattering \cite{Shternin:2008es} and
bulk viscosity due to modified Urca processes based on susceptibilities
for interacting matter \cite{Alford:2010gw}. The dotted curve shows
the exclusion region when using the result for the shear viscosity
from the fit in \cite{Cutler:1990} to the standard low density ($\lesssim n_{0}$)
data from hadron-hadron-scattering given in \cite{Flowers:1976ApJ...206..218F}.
Extrapolating this fit to high densities relevant for neutron stars
overestimates the viscosity compared to its actual, subleading size
\cite{Shternin:2008es}, leading to the smaller exclusion region.
 The dashed curve shows the result when employing the previously
used expression for the bulk viscosity \cite{Cutler:1990,Sawyer:1989dp}
which employs susceptibilities in the approximation of an ideal hadron
gas, see table \ref{tab:strong-parameters}. It is again the insensitivity
to the viscosity parameter $\tilde{V}$ discussed in the previous
subsection that is responsible for the fact that these corrections,
which change the damping time by more than an order of magnitude,
have only such a mild effect on the exclusion region and the minimum
eq. (\ref{eq:NS-minimum-frequency}). It is interesting that the effect
of the interactions is opposite to that of the 2nd order $\Omega$-corrections
\cite{Lindblom:1999yk}. The combined curve formed by the dotted and
the dashed segment is the exclusion region that had previously been
studied in the literature. As can be seen the combined effect of both
corrections is to move the instability region to lower temperatures
so that it extends to roughly $10^{5}$ K. This is relevant for old
stars at low temperatures that are spun up by accretion since the
r-mode becomes unstable already at lower frequencies.

\begin{figure*}
\begin{minipage}[t]{0.5\textwidth}%
\includegraphics{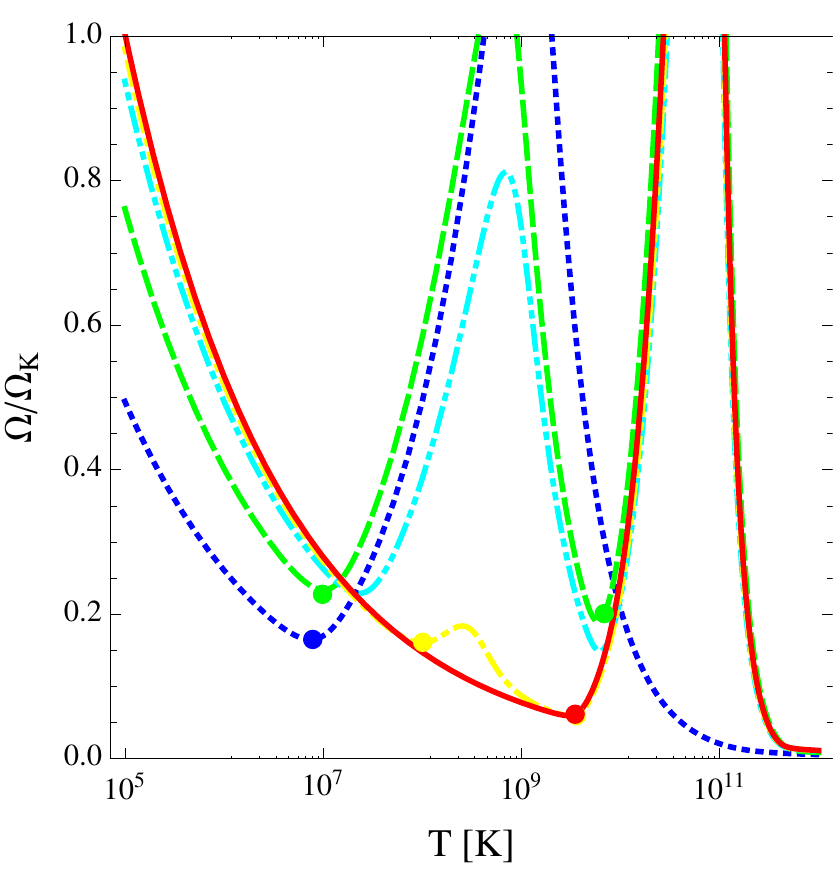}%
\end{minipage}%
\begin{minipage}[t]{0.5\textwidth}%
\includegraphics{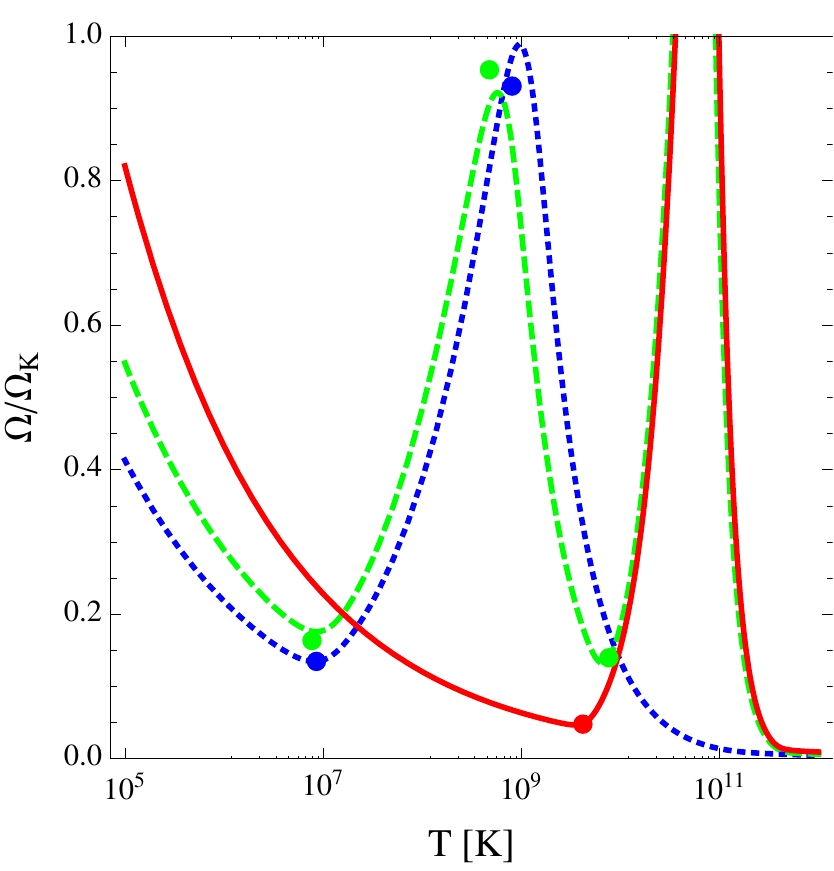}%
\end{minipage}

\caption{\label{fig:instability-regions-all}Instability regions for the different
star models considered in this work. Shown are neutron star models
with APR equation of state (solid), hybrid stars with a small (dot-dashed),
a medium (dot-dot-dashed) and a large quark matter core (dashed) and
strange star models (dotted) with an ideal gas equation of state.
The blobs show again the analytic estimates for the extrema. \emph{Left
panel:} $1.4\, M_{\odot}$ stars. \emph{Right panel:} $2\, M_{\odot}$
stars, for which for the considered equations of state stars with
smaller quark cores could not be found. }

\end{figure*}

The left panel of fig.~\ref{fig:instability-regions-all} compares
the instability regions of the different $1.4\, M_{\odot}$ star models
considered here and also includes the analytic results for the extrema.
The minima in these plots corresponds to the maxima where two damping
mechanisms cross in fig.~\ref{fig:all-damping times} whereas the
maxima in fig.~\ref{fig:instability-regions-all} correspond to the
minima in fig.~\ref{fig:all-damping times}. As can be seen the instability
regions of the hybrid stars interpolate between the neutron star and
the strange star curve as the size of the quark core increases, even
though all these curves are based on rather different equations of
state distinguished by the interaction parameter $c$, see table \ref{tab:star-models}.
The right panel of fig.~\ref{fig:instability-regions-all} shows
the instability regions for the $2\, M_{\odot}$ models discussed
in this work. As has already been observed as a generic feature of
the analytic analysis, the r-mode instability in heavy stars is enhanced
and the boundary of the instability region moves slightly to lower
frequencies. As predicted by the mass dependences of the analytic
expressions this is most pronounced in the vicinity of the maxima
and milder in the vicinity of the minima. The approximate analytic
results for the maxima deviate slightly from the maxima of the numerical
curves. The reason for this is that the uniform density approximation
is not justified in this case due to the strong radial dependence
of the density profiles, see fig.~\ref{fig:density profiles}.

\begin{figure}
\includegraphics{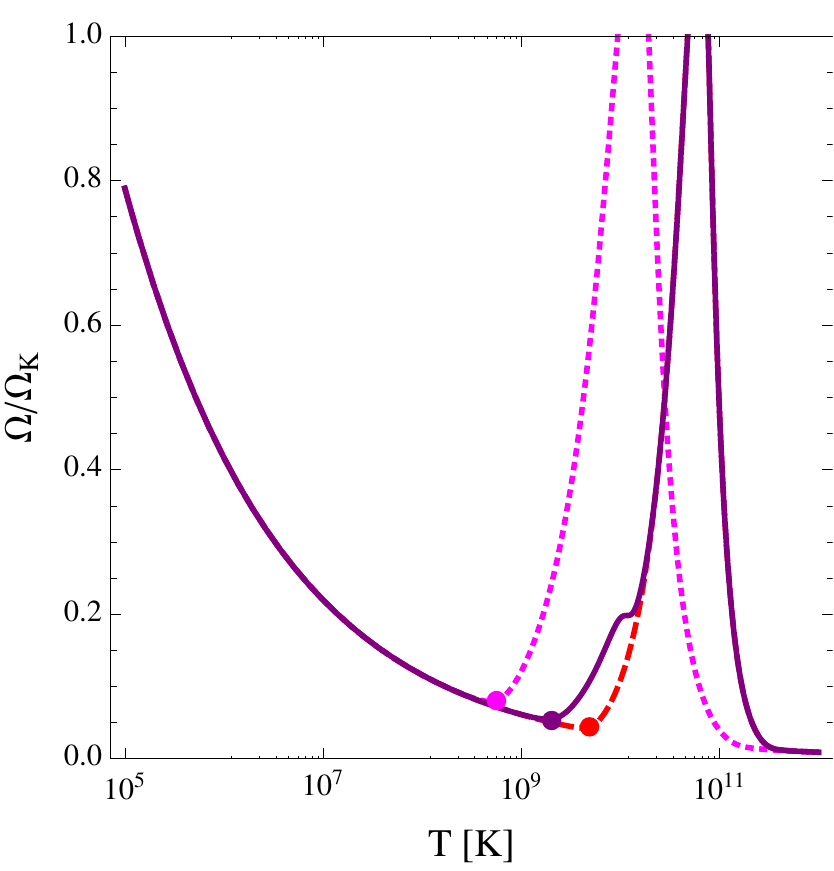}

\caption{\label{fig:direct-Urca-instabilty} The instability region with direct
Urca interactions is shown for the $2.21\, M_{\odot}$ maximum mass
neutron star model (solid curve), where direct Urca processes are
allowed in the inner core but only modified Urca processes in the
layer surrounding it. For comparison the instability region is shown
when direct Urca reactions are artificially suppressed (dashed curve)
as well as when they are artificially allowed in the entire hadronic
core (dotted curve).}

\end{figure}

Fig.~\ref{fig:direct-Urca-instabilty} shows the instability region
for an ultra-heavy neutron star $M\approx2.2\, M_{\odot}$ denoted
by the solid curve. Here, the instability region has grown to the
point that the lower and the upper part are about to merge. For this
star the densities that are reached are high enough that direct Urca
processes are kinematically allowed within an inner core with radius
of roughly $R/2$. The enhanced damping due to direct Urca bulk viscosity
in the inner core leads to a notch at the right side of the instability
region, as also found in \cite{Reisenegger:2003cq}. Yet, the modification
is rather mild in this case: the dashed curve shows the same star
model when direct Urca processes are artificially suppressed. In the
opposite extreme when direct Urca processes are artificially allowed
in the entire core, shown by the dotted curve, the enhanced damping
does lead to a significant change of the instability region. Moreover,
the frequency of the minimum of the instability region increases.
Due to the resonant form of the bulk viscosity, however, the instability
region does not uniformly shrink over the whole temperature range
where bulk viscosity dominates, but the stability window moves to
lower temperatures leaving the r-mode unstable at higher temperatures
where it is otherwise stable in the presence of modified Urca processes.
The temperature scales of the corresponding stability windows agree
with the temperature scales of the resonant maximum of the bulk viscosity,
see \cite{Alford:2010gw}. Direct Urca processes are very sensitive
to the proton fraction of dense matter. Whereas the required fraction
is roughly $14\%$ in the case of the APR equation of state, reached
at relatively high densities $\gtrsim5n_{0}$, this could be different
for other equations of state. In a case where the direct Urca core
is larger but a modified Urca shell is still present so that both
of them have a sizeable volume fraction, their combined damping could
lead to a larger stability window, yet still at parametrically larger
temperatures than the stability window of strange stars.

\begin{figure}
\begin{tabular}{|c|c|}
\hline 
\includegraphics[scale=0.5]{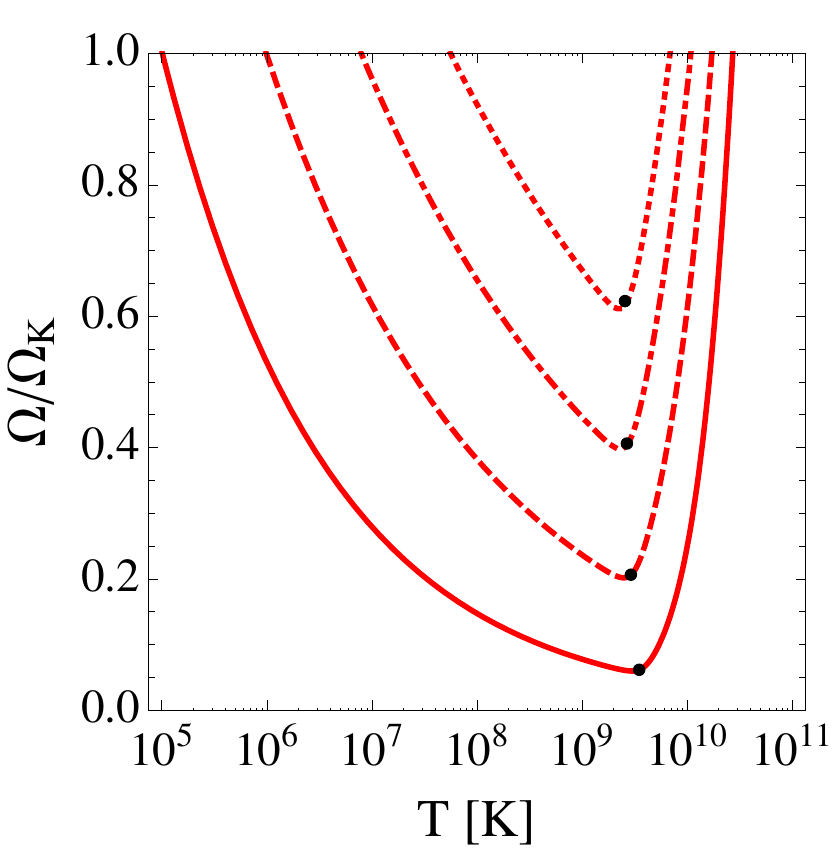} & \includegraphics[scale=0.5]{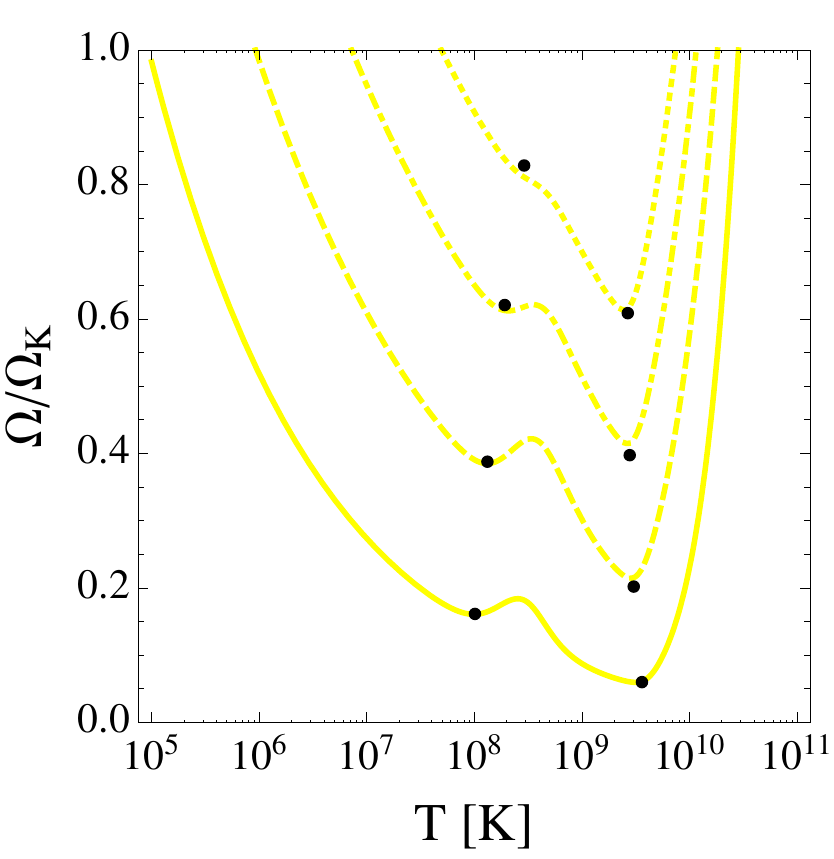}\tabularnewline
\hline
\hline 
\includegraphics[scale=0.5]{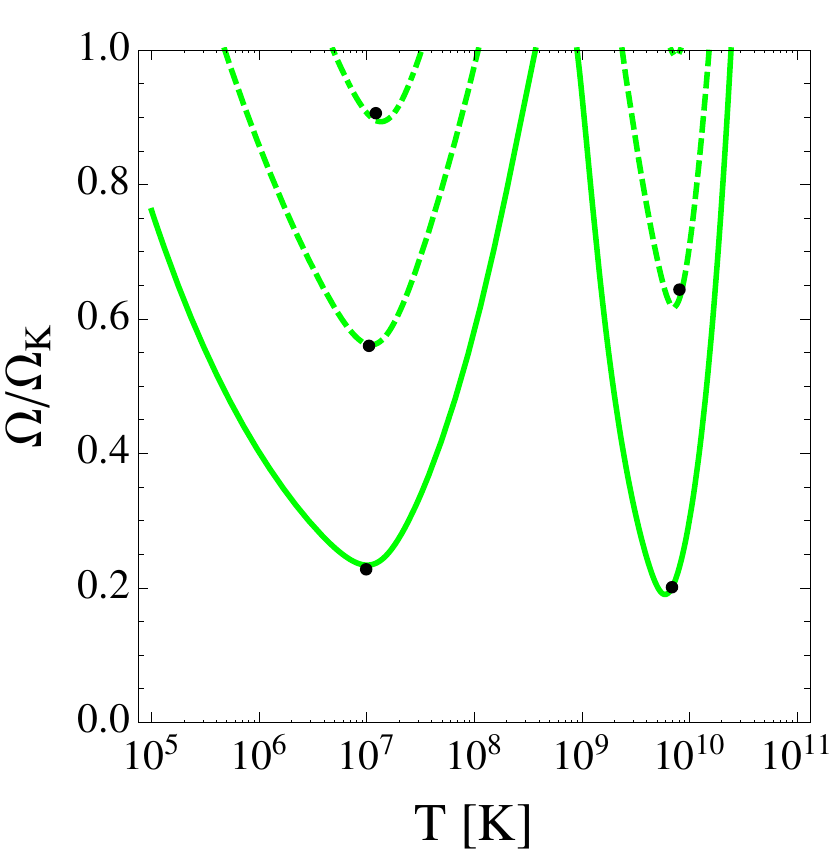} & \includegraphics[scale=0.5]{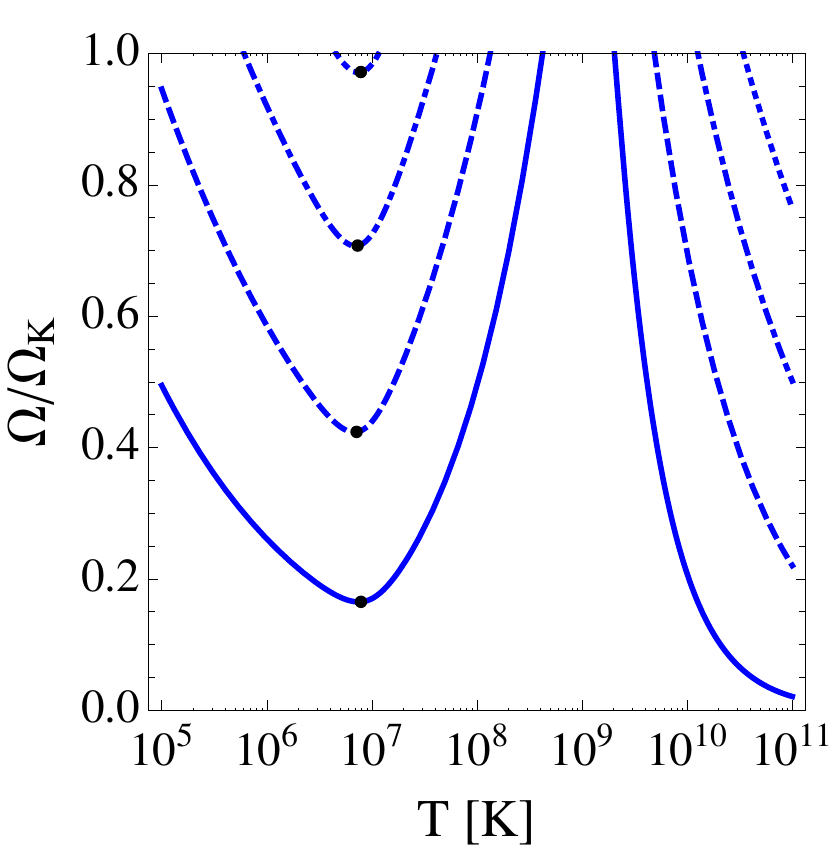}\tabularnewline
\hline
\end{tabular}

\caption{\label{fig:instability-regions-multipoles}Instability regions for
the first four multipole r-modes ($m=2$ to $5$) of the different
$1.4\, M_{\odot}$ star models (top, left: neutron star; top, right:
hybrid star, small quark core; bottom, left: hybrid star, large quark
core; bottom right: strange star). The minima from eqs. (\ref{eq:minimum-frequency})
\& (\ref{eq:minimum-temperature}), the second minima of the large
core hybrid star from eqs. (\ref{eq:second-minimum-frequency}) \&
(\ref{eq:second-minimum-temperature}) and the maxima from eqs. (\ref{eq:maximum-frequency})
\& (\ref{eq:maximum-temperature}) are denoted by the dots. For higher
multipoles the size of the instability regions decreases and they
move to higher rotation frequencies, so that all modes with $m\geq7$
are entirely stable in the physical range of frequencies. In particular
in the cases of neutron and hybrid stars the right boundaries of the
instability regions of the higher multipoles are very close to the
fundamental $m=2$ mode, so that several modes can easily be excited
if the evolution significantly enters the fundamental instability
region. }

\end{figure}

Fig.~\ref{fig:instability-regions-multipoles} shows the result for
the higher multipoles for the different $1.4\, M_{\odot}$ star models.
As can be seen, the right boundaries of the instability regions for
the higher multipoles are extremely close to the boundary of the fundamental
$m=2$ mode. Correspondingly these modes could easily be excited once
the evolution of a star enters the instability region. Each mode that
is triggered would lead to a further enhancement of the spindown of
the star and could change the evolution. This could be particularly
relevant for young neutron stars that enter the instability region
at high temperatures when the cooling is still fast but there are
no strong reheating mechanisms which could prevent the star from substantially
penetrating the instability region.

\begin{figure}
\includegraphics{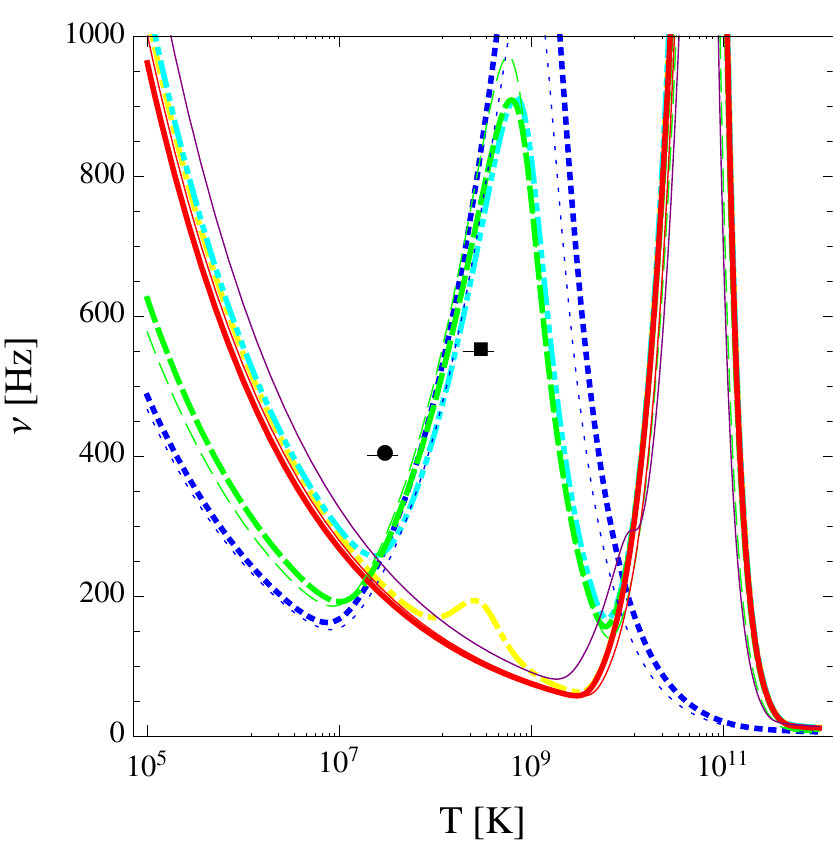}

\caption{\label{fig:data-comparisson}Comparison of the instability regions
in absolute frequencies for the different star models considered in
this work with the two low mass X-ray binaries Aql X-1 (filled square)
and SAX J1808.4-3658 (open circle). The horizontal bar gives a partial
measure for the error within the model computations which should be
larger due to uncontrolled assumptions that are not considered in
its size.}

\end{figure}

Fig.~\ref{fig:data-comparisson} finally compares our numeric results
to astrophysical data. Although the rotation frequencies of pulsars
are known to high accuracy, the relevant core temperatures are not
known for most pulsars and involve large uncertainties even for those
stars where estimates are available. The reason is that the temperature
of the surface is indirectly inferred from the data and it is necessary
to abstract from it the core temperature via models. Such an analysis
has been performed for two low mass X-ray binary systems Aql X-1 and
SAX J1808.4-3658 \cite{Brown:2002rf}, where the pulsars accrete matter
from companion stars. The horizontal lines give an optimistic estimate
for the uncertainty of the temperature. As can be seen these data
points are well within the instability region of neutron stars. This
holds in particular for the faster one of the two pulsars, Aql X-1,
which is right in the middle. Our analysis shows this statement cannot
be undermined by the unknown equation of state. So if r-modes would
spin down stars so fast that the stars would leave the unstable region
on time scales short compared to astrophysical ones these stars could
not be pure neutron stars and would require some form of enhanced
damping. Yet, in the case of strange stars and hybrid stars with a
sufficiently large quark core, the data points are close to the boundary
of the corresponding stability window and would be compatible in such
a scenario%
\footnote{A potential problem with such an explanation of the data would be
the observation of a very fast and at the same time very cold pulsar,
since, after the accretion stops, stars within the stability window
would re-enter the instability region due to cooling and correspondingly
would have to spin down to a fraction of their Kepler frequency.%
}. As our results show, hybrid stars with a small quark core or neutron
stars with enhanced hadronic direct Urca damping, in contrast, would
not be sufficient. 

\section{Conclusions}

Using general expressions for the viscosities of dense matter we have
derived semi-analytic results for the damping times of small amplitude
r-mode oscillations and the boundary of the instability region. Our
results show that the boundary of the instability region and in particular
its minimum, which determines to what extent r-modes can spin down
a fast star, are extremely insensitive to the quantitative details
of the microscopic interactions that induce viscous damping. However,
the instability regions can nevertheless effectively discriminate
between qualitatively different classes of stars. In particular strange
stars and hybrid stars with sufficiently large quark cores feature
a stability window that cannot be reproduced with standard neutron
stars without some admixture of exotic matter that provides enhanced
damping. We find that the presence of some form of exotic matter does
not automatically lead to a stability window since the instability
region of hybrid stars with a sufficiently small quark core is almost
indistinguishable from that of a neutron star. Similarly, the presence
of neutron matter with direct Urca interactions will in most cases
not considerably change the instability region. However, due to the
demonstrated insensitivity of the instability regions to quantitative
microscopic details, the clear determination of a very fast pulsar
with $\Omega>300$ Hz in the temperature range around $10^{9}$ K
could provide a convincing signature for some exotic form of matter.
What remains to be shown in order to transform this into a strict
signature is that the crust does not dominate the damping, and that
r-modes do not saturate at amplitudes that are so small that the spindown
takes billions of years in which case the instability region would
not really present a no-go area. The second point requires a thorough
understanding of the dynamical evolution of compact stars and a step
towards this goal is taken in a companion paper \cite{Alford:2011pi}
where we show that the large amplitude behavior of the bulk viscosity
can saturate r-modes at amplitudes that are large enough for a fast
spindown.
\begin{acknowledgments}
We thank Nils Andersson, Greg Comer, Brynmor Haskell, Prashant Jaikumar,
Andreas Reisenegger and Andrew Steiner for helpful discussions. This
research was supported in part by the Offices of Nuclear Physics and
High Energy Physics of the U.S. Department of Energy under contracts
\#DE-FG02-91ER40628, \#DE-FG02-05ER41375. 
\end{acknowledgments}
\appendix

\section{Analytic form for the R-mode profile of constant density stars\label{sec:Analytic-r-mode}}

In this appendix we give an analytic solution for the density fluctuation
profile of the r-mode to leading order in $\Omega$ for a star of
constant density. The part of the leading order density fluctuation
eq. (\ref{eq:r-profile}) that is generally not analytic is given
by the change in the gravitational potential $\delta\Phi_{0}$ whose
radial part $\delta\Phi_{0}\!\left(r\right)$ is defined by

\[
\delta\Phi_{0}\!\left(\vec{r}\right)=\sqrt{\frac{m}{\pi\left(m\!+\!1\right)^{3}\left(2m\!+\!1\right)!}}\alpha\delta\Phi_{0}\!\left(r\right)P_{m+1}^{m}\!\left(\cos\theta\right)\mathrm{e}^{im\varphi}\]
It fulfills the differential equation \cite{Lindblom:1999yk}

\begin{align*}
 & \frac{d^{2}\delta\Phi_{0}}{dr^{2}}+\frac{2}{r}\frac{d\delta\Phi_{0}}{dr}+\left(4\pi G\rho\frac{d\rho}{dp}-\frac{\left(m+1\right)\left(m+2\right)}{r^{2}}\right)\delta\Phi_{0}\\
 & \qquad=-4\pi G\rho\frac{d\rho}{dp}\left(\frac{r}{R}\right)^{m+1}\end{align*}
with boundary conditions $\delta\Phi_{0}\!\left(0\right)=0$ and

\[
\left.\frac{d\delta\Phi_{0}\!\left(r\right)}{dr}\right|_{r=R}=-\left(\frac{1}{2}+\sqrt{\frac{1}{4}+\left(m+1\right)\left(m+2\right)}\right)\frac{\delta\Phi_{0}\!\left(R\right)}{R}\]
Here we specialize to the fundamental $m=2$ r-mode where an analytic
solution is possible for the idealized case $\rho=const.$ and $d\rho/dp=const.$
which is approximately realized for strange stars. The solution obtained
with a computer algebra system reads

\begin{align*}
 & \delta\Phi_{0}\!\left(r\right)=-\frac{r^{3}}{R^{3}}\\
 & +\frac{7LR^{2}\left(\left(r^{3}-15L^{2}r\right)\cos\!\left(\frac{r}{L}\right)+3L\left(5L^{2}-2r^{2}\right)\sin\!\left(\frac{r}{L}\right)\right)}{r^{4}\left(\left(3L^{2}-R^{2}\right)\sin\!\left(\frac{R}{L}\right)-3LR\cos\!\left(\frac{R}{L}\right)\right)}\end{align*}
in terms of the intrinsic length scale

\[
L\equiv\frac{1}{\sqrt{4\pi G\rho\frac{\delta\rho}{\delta p}}}\]
Unfortunately the above expression is rather ill behaved due to strong
cancellations and not suitable for a direct evaluation. However, employing
the series representations of the trigonometric functions it is possible
to transform it into an alternative form \[
\delta\Phi_{0}\!\left(r\right)=g\left(\frac{r}{L},\frac{R}{L}\right)\frac{r^{3}}{R^{3}}\]
in terms of hypergeometric functions

\[
g\!\left(x,y\right)\equiv\frac{_{0}F_{2}\!\left(\frac{1}{2},\frac{7}{2}+1;-\frac{x^{2}}{8}\right)}{_{0}F_{2}\!\left(\frac{1}{2},\frac{5}{2}+1;-\frac{y^{2}}{8}\right)}-1\approx-\frac{x^{2}}{18}+\frac{y^{2}}{14}+\cdots\]
Correspondingly, $\left|g\right|<0.07$ over the entire parameter
range so that in view of the large uncertainties inherent in an r-mode
analysis the gravitational potential term in eq. (\ref{eq:r-profile})
can be neglected leaving a simple analytic expression for the r-mode
profile of strange stars with approximately constant density.

\section{General semi-analytic expressions for the boundary of the instability
region\label{sec:Semi-analytic-boundary}}

In this appendix we give the general semi-analytic expressions for
the boundary of the instability region that are valid for arbitrary
multipoles and can be applied to stars involving several shells. The
only non-analytic input in these expressions enters via the radial
integral constants $\tilde{J}$, $\tilde{S}$, $\tilde{V}$ and $\tilde{W}$
arising in the gravitational time scale, the shear viscosity time
scale as well as the low and high temperature limit of the bulk viscosity
time scale. In general these constants, which are given for the star
models discussed in this work and the normalization scales $\Lambda_{QCD}=1$
GeV and $\Lambda_{EW}=100$ GeV in table \ref{tab:parameter-extrema-values},
have to be computed numerically. However, since the dependence of
many of the following results on these constants is surprisingly weak,
these expressions are often essentially analytic.

On the low temperature side of the instability region the shear viscosity
dominates and neglecting the bulk viscosity yields the analytic result
for the boundary in this region

\begin{align}
\Omega\!\left(T\right) & \xrightarrow[T\ll T_{min}]{}\left(\frac{\left(2m+1\right)\left(\left(2m+1\right)!!\right)^{2}\left(m+1\right)^{2m+2}}{32\pi\left(m+2\right)^{2m+2}\left(m-1\right)^{2m-1}}\right.\nonumber \\
 & \qquad\qquad\qquad\cdot\left.\frac{\tilde{S}_{m}\Lambda_{QCD}^{3+\sigma}}{\tilde{J}_{m}^{2}GM^{2}R^{2m-1}T^{\sigma}}\right)^{\frac{1}{2m+2}}\end{align}
In the case of a hybrid star with two or more distinct shells one
of them generally dominates and the above expression applies using
only the contribution $\tilde{S}$ of the dominant shell as well as
the corresponding exponent $\sigma$. The dominant contribution to
the inverse shear viscosity damping time arises from the hadronic
shell. Since the shear damping time increases monotonically with temperature
whereas the bulk viscosity decreases in this regime monotonically,
the minimum of the instability region is taken at the temperature
where $\tau_{S}=\tau_{B}$. The minimum frequency is then obtained
from $1/\tau_{G}+2/\tau_{B}=0$ and taking into account that at low
frequency $\kappa\approx\kappa_{0}$ it yields

\begin{align}
 & \Omega_{min}\approx\left(\left(\frac{m\left(m+1\right)^{2m-1}\left(\left(2m+1\right)!!\right)^{2}}{4\pi\left(2m+3\right)\left(m+2\right)^{2m+2}\left(m-1\right)^{2m}}\right)^{\sigma}\right.\label{eq:minimum-frequency}\\
 & \quad\;\cdot\left(\frac{\left(\left(2m\!+\!1\right)!!\right)^{2}\left(2m\!+\!1\right)\left(m\!+\!1\right)^{2m+2}}{16\pi\left(m-1\right)^{2m-1}\left(m+2\right)^{2m+2}}\right)^{\delta}\nonumber \\
 & \cdot\left.\frac{\tilde{V}_{m}^{\sigma}\tilde{S}_{m}^{\delta}\Lambda_{QCD}^{3\delta+9\sigma}}{\tilde{J}_{m}^{2\left(\delta+\sigma\right)}\Lambda_{EW}^{4\sigma}G^{\delta+\sigma}R^{2m\left(\delta+\sigma\right)-\delta-5\sigma}M^{2\left(\delta+\sigma\right)}}\right)^{\frac{1}{2m\left(\delta+\sigma\right)+2\delta}}\nonumber \end{align}
whereas the corresponding minimum temperature is

\begin{align}
T_{min} & \approx\left(\frac{\left(2m+3\right)\left(2m+1\right)\left(m+1\right)^{3}\left(m-1\right)}{4m}\right.\nonumber \\
 & \qquad\qquad\left.\cdot\frac{\tilde{S}_{m}\Lambda_{EW}^{4}\Lambda_{QCD}^{\delta+\sigma-6}}{\tilde{V}_{m}R^{4}}\right)^{\frac{1}{\delta+\sigma}}\Omega_{min}^{-\frac{2}{\delta+\sigma}}\label{eq:minimum-temperature}\end{align}
Above the minimum the bulk viscosity dominates and increases until
it reaches a maximum. Within a temperature range in between these
two extrema shear viscosity can be neglected and the low temperature
approximation to the bulk viscosity can be applied. In the general
case the equation $1/\tau_{B}+1/\tau_{G}=0$ for the boundary cannot
be solved analytically as a function of $T$ due to the non-linear
second order terms in $\Omega$. However, it is possible to solve
the equation analytically as a function of $\Omega$ which provides
an equally valid parameterization of the segment in between the two
extrema 

\begin{align}
T\!\left(\Omega\right) & \xrightarrow[\Omega\ll\Omega_{max}]{\Omega_{min}\ll\Omega}\left(\!\frac{2\pi\left(2m\!+\!3\right)\!\left(m\!+\!2\right)^{2m\!+\!2}\!\left(m\!-\!1\right)^{2m}\!\kappa^{2}}{m\left(m+1\right)^{2m-3}\left(\left(2m+1\right)!!\right)^{2}}\right.\nonumber \\
 & \qquad\qquad\qquad\left.\cdot\frac{\tilde{J_{m}}^{2}\Lambda_{EW}^{4}GM^{2}\Omega^{2m}}{\tilde{V}_{m}\Lambda_{QCD}^{9-\delta}R^{5-2m}}\right)^{\frac{1}{\delta}}\end{align}
Since there is no analytic expression for the bulk viscosity around
the maximum only for the special case of constant density stars a
result is possible. In this case the maximum frequency is

\begin{align}
\Omega_{max} & \xrightarrow[\mathrm{dens.}]{\mathrm{unif.}}\left(\frac{2\pi m\left(m+1\right)^{2m-2}\left(2m+3\right)\left(\left(2m+1\right)!!\right)^{2}}{9\left(2m+5\right)\left(m+2\right)^{2m+2}\left(m-1\right)^{2m}}\right.\nonumber \\
 & \qquad\qquad\qquad\left.\cdot\frac{\bar{A}^{2}\bar{C}^{2}\left(R_{o}^{3}-R_{i}^{3}\right)}{G\bar{B}M^{2}R^{2m-2}}\right)^{\frac{1}{2m-1}}\label{eq:maximum-frequency}\end{align}
where $R_{i}$ and $R_{o}$ are the inner and outer radii of the dominant
shell ($0$ and $R$ for a homogeneous star) and bars denote average
quantities over the shell in case the density is only approximately
constant. The corresponding maximum temperature is

\begin{equation}
T_{max}\xrightarrow[\mathrm{density}]{\mathrm{uniform}}\left(\frac{2\Omega_{max}}{\left(m+1\right)\bar{\tilde{\Gamma}}\bar{B}}\right)^{\frac{1}{\delta}}\label{eq:maximum-temperature}\end{equation}
In general the constant density approximation is only valid for quark
matter and the corresponding results in this case are given in the
main text. However, for neutron stars the corresponding maximum is
in general above the Kepler frequency anyway and therefore not physically
interesting. Far above the temperature where the maximum is located
the high temperature approximation to the bulk viscosity is valid
and the boundary $1/\tau_{B}+1/\tau_{G}=0$ condition yields here

\begin{align}
\Omega\!\left(T\right) & \xrightarrow[T\gg T_{max}]{}\left(\frac{m\left(m+1\right)^{2m-3}\left(\left(2m+1\right)!!\right)^{2}}{2\pi\left(2m+3\right)\left(m+2\right)^{2m+2}\left(m-1\right)^{2m}}\right.\nonumber \\
 & \qquad\qquad\quad\left.\cdot\frac{\tilde{W}_{m}\Lambda_{EW}^{4}\Lambda_{QCD}^{\delta-1}R^{5-2m}}{\tilde{J}_{m}^{2}GM^{2}T^{\delta}}\right)^{\frac{1}{2m-2}}\end{align}
For homogeneous stars the present approximations cover nearly the
entire boundary of the instability region. If the star has more than
one layer, there may be a second minimum. Depending on the size of
the layers here it can be either the bulk viscosity of the second
layer that equals the dominant shear viscosity at the minimum, in
which case the above expressions hold, or the bulk viscosities of
the different layers are identical $\tau_{B}^{\left(i\right)}=\tau_{B}^{\left(o\right)}$
which yields analogously

\begin{align}
 & \Omega_{min}^{\left(B\right)}=\left(\left(\frac{m\left(m+1\right)^{2m-3}\left(\left(2m+1\right)!!\right)^{2}}{\pi\left(2m+3\right)\left(m-1\right)^{2m}\left(m+2\right)^{2m+2}}\right)^{\delta^{\left(o\right)}+\delta^{\left(i\right)}}\right.\nonumber \\
 & \cdot\left(\frac{m+1}{2}\right)^{2\delta^{\left(i\right)}}\frac{\Lambda_{QCD}^{9\delta^{\left(i\right)}-\delta^{\left(o\right)}}\left(\tilde{V}_{m}^{\left(o\right)}\right)^{\delta^{\left(i\right)}}\left(\tilde{W}_{m}^{\left(i\right)}\right)^{\delta^{\left(o\right)}}}{\Lambda_{EW}^{4\left(\delta^{\left(i\right)}-\delta^{\left(o\right)}\right)}G^{\left(\delta^{\left(o\right)}+\delta^{\left(i\right)}\right)}\tilde{J}_{m}^{2\left(\delta^{\left(o\right)}+\delta^{\left(i\right)}\right)}}\nonumber \\
 & \qquad\cdot\left.\frac{R^{\left(5-2m\right)\left(\delta^{\left(o\right)}+\delta^{\left(i\right)}\right)}}{M^{2\left(\delta^{\left(o\right)}+\delta^{\left(i\right)}\right)}}\right)^{\frac{1}{2m\left(\delta^{\left(o\right)}+\delta^{\left(i\right)}\right)-2\delta^{\left(o\right)}}}\label{eq:second-minimum-frequency}\end{align}
and the corresponding temperature is

\begin{align}
T_{min}^{\left(B\right)} & =\left(\frac{4}{\left(m+1\right)^{2}}\right.\label{eq:second-minimum-temperature}\\
 & \left.\cdot\frac{W_{m}^{\left(i\right)}\Lambda_{EW}^{8}\Lambda_{QCD}^{\delta^{\left(o\right)}+\delta^{\left(i\right)}-10}}{V_{m}^{\left(o\right)}}\right)^{\frac{1}{\delta^{\left(o\right)}+\delta^{\left(i\right)}}}\left(\Omega_{min}^{\left(B\right)}\right)^{\frac{2}{\delta^{\left(o\right)}+\delta^{\left(i\right)}}}\nonumber \end{align}

\bibliographystyle{JHEP_MGA}
\bibliography{cs}

\end{document}